\documentclass[fleqn,usenatbib]{mnras}
\usepackage{amsmath}    
\usepackage{newtxtext,newtxmath}
\usepackage[T1]{fontenc}
\usepackage{ae,aecompl}
\usepackage{graphicx}	
\usepackage{geometry}
\usepackage{float}
\usepackage{caption}
\usepackage{subcaption}
\usepackage{dblfloatfix}
\usepackage{url}
\usepackage{xcolor,colortbl}

\newcommand{\hmpc}{\,h^{-1}\,{\rm Mpc}}

\newcommand{\hmsun}{\,h^{-1}\,{\rm M}_\odot}
\newcommand{\vvir}{{V_{\rm vir}}}

\newcommand{\vmax}{{V_{\rm max}}}
\newcommand{\vpeak}{{V_{\rm peak}}}
\newcommand{\ahalf}{{a_{\rm 0.5}}}

\newcommand{\Msun}{{\rm M}_\odot}

\title{Conditional Colour-Magnitude Distribution of Central Galaxies in Galaxy Formation Models}

\author[X. Xu et al.]{
Xiaoju Xu,$^{1}$\thanks{E-mail: xiaoju@sjtu.edu.cn}
Zheng Zheng,$^{2}$\thanks{E-mail: zhengzheng@astro.utah.edu}
and Qi Guo$^{3,4}$\\
$^{1}$Department of Astronomy, School of Physics and Astronomy, Shanghai Jiao Tong University,
Shanghai 200240, China\\
$^{2}$Department of Physics and Astronomy, University of Utah, 115 South 1400 East, Salt Lake City, UT 84112, USA\\
$^{3}$Key Laboratory for Computational Astrophysics, National Astronomical Observatories, Chinese
Academy of Sciences, Beijing 100012, China\\
$^{4}$School of Astronomy and Space Science, University of Chinese Academy of Sciences, Beijing 100049, China}

\date{Accepted XXX. Received YYY; in original form ZZZ}

\pubyear{2022}

\begin{document}
\label{firstpage}
\pagerange{\pageref{firstpage}--\pageref{lastpage}}
\maketitle

\defcitealias{XuHaojie2018}{Xu18}

\begin{abstract}

We investigate the conditional colour-magnitude distribution (CCMD), namely the colour-magnitude distribution at fixed halo mass, of the central galaxies in semi-analytic galaxy formation model (SAM) and hydrodynamic simulations. We analyse the CCMD of central galaxies in each halo mass bin with the Gaussian mixture model and find that it can be decomposed into red and blue components nearly orthogonal to each other, a red component narrow in colour and extended in magnitude and a blue component narrow in magnitude and extended in colour. We focus on the SAM galaxies to explore the origin of the CCMD components by studying the relation between central galaxy colour and halo or galaxy properties. Central galaxy colour is correlated with halo assembly properties for low mass haloes and independent of them for high mass haloes. Galaxy properties such as central supermassive black hole mass, cold gas mass, and gas specific angular momentum can all impact central galaxy colour. These results are corroborated by an alternative machine learning analysis in which we attempt to predict central galaxy colour with halo and galaxy properties. We find that the prediction for colours of central galaxies can be significantly improved using both halo and galaxy properties as input compared to using halo properties alone. With the halo and galaxy properties considered here, we find that subtle discrepancies remain between predicted and original colour distribution for low mass haloes and that no significant determining properties are identified in massive haloes, suggesting modulations by additional stochastic processes in galaxy formation.
  \end{abstract}


\section{Introduction}
\label{sec:intro}

Observationally, magnitude and colour are fundamental galaxy properties that can be directly measured. It is well known that galaxy colour exhibits bimodality in the colour-magnitude diagram (CMD) with a red sequence and a blue cloud \citep{Strateva2001,Blanton2003,Baldry2004,Balogh2004,Taylor2009}. The blue cloud consists of galaxies forming stars actively, whereas the red sequence presents due to the quenching of star formation in those galaxies. Closely relating to physical properties such as stellar mass and specific star formation rate (sSFR), magnitude and colour are important indicators of galaxy evolution status and key galaxy properties to be recovered with galaxy formation models.

Galaxy formation and evolution can be modeled with semi-analytic models (SAM, \citealt{DeLucia2004,Croton2006,Guo2011,Guo2013,Cora2018}) that implement physical processes using haloes from $N$-body simulations and with hydrodynamic simulations that evolve baryonic particles together with dark matter particles \citep{Vogelsberger2014,Schaye2015,Dave2019,Nelson2019}. These models can provide detailed galaxy properties including direct observables and physical properties at each snapshot of evolution and predict galaxy clustering consistent with observation. However, they are usually computationally expensive, especially the hydrodynamic simulations in large cosmological volumes. 

Since galaxies form and evolve in dark matter haloes \citep{White1978}, empirical models such as halo occupation distribution (HOD, \citealt{Peacock2000,Berlind02,Zheng2005,Zheng09,Zheng07}) and conditional luminosity function (CLF, \citealt{Bosch03,Bosch13,Cooray06,Yang03,Yang08}) are also widely implemented for modelling galaxy clustering by connecting halo occupation or galaxy luminosity distribution to halo. Focusing on subhalo catalogue of $N$-body simulations, subhalo abundance matching (SHAM) can also model galaxy clustering through linking galaxy stellar mass or luminosity to subhalo mass or other mass
indicators \citep{Conroy2006,Guo2010,Reddick2013,Zentner2014,Chaves2016,Guo2016}. All these models assume that galaxy occupation number or galaxy property only depends on halo mass.  

However, galaxy properties such as stellar mass are found to depend not only on halo mass but also on other halo properties such as halo age, concentration, and environment in SAM or hydrodynamic simulations \citep{Artale2018,Zehavi2018,Bose2019,Martizzi2020,Xu2020,Xu2021a}. Modified empirical models are built to include these secondary effects, such as the decorated HOD \citep{Hearin2016} and the environment-dependent HOD \citep{McEwen2018,Wibking2019,Hadzhiyska2021,Xu2021a}. Traditional SHAM can also be modified to incorporate dependencies of stellar mass on secondary subhalo properties \citep{Contreras2021}. These models can reproduce galaxy clustering of stellar-mass threshold samples to a high precision.  

Besides clustering of luminosity-selected sample, clustering of colour-selected sample and joint luminosity-colour sample are also measured in observations \citep[e.g.][]{Zehavi2005,Zehavi2011,Coil2008}. For modelling these samples, it is required to further improve the empirical models connecting galaxy colour to halo or subhalo properties. Efforts have been made by modifying the SHAM to assign galaxy colour or SFR in subhaloes in addition to stellar mass \citep[e.g.][]{Hearin2013,Hearin2014,Contreras2021,Favole2022}. 

For modelling more detailed colour-dependent clustering, it is crucial to recover the colour distribution in observations (i.e. the colour bimodality). \citet[][hereafter Xu18]{XuHaojie2018} introduce the conditional color-magnitude distribution (CCMD) model, which considers the CMD to be consist of four components at fixed halo mass bins, two for centrals and two for satellites. Each central component is described by a 2D Gaussian distribution, and each satellite component is described by a Schechter-like function for luminosity and Gaussian distribution for colour. Fitting the observed clustering in fine luminosity-colour bins of galaxies from the Sloan Digital Sky Survey (SDSS, \citealt{York2000}), they find that the two Gaussian components for central galaxies, named the pseudo-red and pseudo-blue central components, are orthogonal --- the former expands a narrow colour range and wide luminosity range, whereas the latter expands a wide colour range and a narrow luminosity range. They demonstrate that the CCMD model can recover the joint dependencies of observed clustering on luminosity and colour reasonably well. 

Similar to the red sequence in CMD, the pseudo-red component in CCMD also mainly consists of quenched galaxies. Plenty of studies investigated the reason for quenching in literature, among which feedback and environment quenching are widely explored. Internally, stellar and active galactic nuclei (AGN) feedback causes gas outflow and gas temperature increase, which could slow down the star formation activity \citep{Bower2006,Croton2006,vandeVoort2016,Montero2019,Davies2021}. Recently, \citet{Cui2021} investigate the origin of colour bimodality in SIMBA simulation and find that the Jet-mode AGN feedback and X-ray AGN feedback play major roles in galaxy quenching. Externally, star formation can also be shut down due to the environment such as gas stripping by massive neighbors or lack of gas inflow \citep{Woo2013,Malavasi2017,Lemaux2019,PintosCastro2019,Sin2019,Gouin2020}. 

The empirically inferred CCMD in \citetalias{XuHaojie2018} can form crucial tests to galaxy formation models. Conversely it is important to understand the physical origin of the different CCMD components based on galaxy formation models. In this paper, we explore the existence and origin of the CCMD components in SAM and hydrodynamic simulations as well as the dependence of galaxy colour on halo properties. For simplicity, we only focus on central galaxies, since centrals and satellites may experience different evolution paths and correlate with halo or environment differently. At fixed halo mass, we explore the possibility to decompose the CCMD from galaxy formation models into two 2D Gaussian components using the Gaussian mixture modelling (GMM). 
We first explore the \citet{Guo2011} SAM galaxy sample and present the GMM decomposition. Since the red and blue components can be differentiated by colour, we then investigate the relations between galaxy colour and halo or galaxy properties. Additionally, we implement machine learning methods such as random forest (RF) to learn these relations to gain more insights on the importance of halo or galaxy properties for galaxy colour. We also perform the GMM analysis with the Illustris simulation \citep{Vogelsberger2014} and the TNG simulation \citep{Nelson2019} and find that the CCMD red and blue components show up in a wider halo mass range for the latter. This study can provide us a better insight of the relation between galaxy properties and halo properties and can help explain galaxy clustering and understand galaxy evolution. 

This paper is organized as follows. In section ~\ref{sec:method}, we introduce the SAM galaxy sample and the hydrodynamic simulations we used, as well as the halo properties we focus on. In section~\ref{sec:samCCMD} we present the CCMD components found in SAM and investigate the relation between central colour and halo or galaxy properties. Additionally, we adopt machine learning to further explore this relation in section~\ref{sec:RF}. Then in section~\ref{sec:hydro_ccmd}, we present the CCMD found in Illustris and TNG. Finally, We summarise the results in section~\ref{sec:summary}.

For simplicity, in this paper we call the CCMD components as red/blue, instead of pseduo-red/pseudo-blue (\citetalias{XuHaojie2018}), unless otherwise noted to avoid confusion.

\section{Data and method}
\label{sec:method}

\subsection{Semi-analytic galaxy formation model and halo properties}
\label{subsec:guo2011}

We first explore CCMD predicted in the galaxy sample of \citet{Guo2011} SAM implemented on the Millennium simulation \citep{Springel2005}. Running in a cosmological volume with a box length of 500 $h^{-1}{\rm Mpc}$ (comoving), the Millennium simulation evolves $2160^3$ dark matter particles with mass $8.6\times10^8 \hmsun$ from $z=127$ to $z=0$. Dark matter haloes and subhaloes are then identified by \texttt{FOF} and \texttt{SUBFIND} algorithms at each snapshot respectively \citep{Davis1985,Springel2001}. A spatially-flat $\Lambda$ cold dark matter ($\Lambda$CDM) cosmological model and the first-year {\it WMAP} data \citep{Spergel2003} are adopted, with parameters $\Omega_{\rm m}=0.25$, $\Omega_{\rm b}=0.045$, $h=0.73$, and $n_s$=1, and $\sigma_8=0.9$. We study the $z=0$ galaxy population. We only use the haloes with mass above $10^{11.4} \hmsun$, since the red component of central galaxies is not clear below this mass.

We focus on both halo internal properties and environmental properties investigating their connection with the galaxy CCMD components. The halo internal properties we used are: 
\begin{itemize}
    \item[(1)] $M_{\rm h}$, the halo mass, specifically $M_{\rm 200c}$ such that the mean density of a halo is 200 times the critical density of the background universe, in units of $\hmsun$;
    \item[(2)] $c$, the halo concentration. We use the definition of $c=\vmax/\vvir$ as a proxy, where $\vmax$ is the maximum circular velocity of the halo and $\vvir$ is the virial velocity;
    \item[(3)] $\vmax$, the maximum circular velocity of the halo;
    \item[(4)] $\vpeak$, the peak value of $\vmax$ over the accretion history of the halo;
    \item[(5)] $a_{\rm 0.5}$, the scale factor at which the halo reaches for the first time 0.5 of its final mass;
    \item[(6)] $a_{\rm 0.8}$, the scale factor at which the halo reaches for the first time 0.8 of its final mass;
    \item[(7)] $a_{\rm vpeak}$, the scale factor when $\vmax (a_{\rm vpeak}) = \vpeak$;
    \item[(8)] $j_{\rm halo}$, the specific angular momentum of the halo (angular momentum per unit mass), in units of $\rm km\, s^{-1} Mpc\, M_\odot^{-1}$;
    \item[(9)] $\dot {M}_{\rm h}$, the average mass accretion rate of the halo from $z=0.2$ to $z=0$ (roughly one dynamical time), in units of $\hmsun {\rm yr}^{-1}$;
    \item[(10)] $\dot {M}_{\rm h}/{M}_{\rm h}$, the specific average mass accretion rate, in units of ${\rm Gyr}^{-1}$;
    \item[(11)] $N_{\rm merg}$, the total number of major mergers on the main branch of halo merger tree, where a major merger is defined as a merger with the mass ratio of the two progenitors exceeding 1/3;
    \item[(12)] $a_{\rm first}$, the scale factor of the first major merger on the main branch of the halo merger tree;
    \item[(13)] $a_{\rm last}$, the scale factor of the last major merger on the main branch of the halo merger tree.   
\end{itemize}

The environmental properties we considered include:
\begin{itemize}
    \item[(1)] $\delta_{1.25}$, the dark matter density smoothed by a Gaussian filter with smoothing scale (standard deviation) of $1.25 \hmpc$; 
    \item[(2)] $\delta_{2.5}$, the dark matter density smoothed by a Gaussian filter with smoothing scale of $2.5 \hmpc$;
    \item[(3)] $\alpha_{\rm 0.3,1.25}$, the tidal anisotropy \citep{Paranjape2018} measured with a Gaussian smoothing scale of $1.25 \hmpc$, defined as
      \begin{equation}
      \label{eq:alpha}  
      \alpha_{\rm n,R}\equiv\left. \sqrt{q_R^2}\right/(1+\delta_R)^{n},
      \end{equation}
      where $n$ is the normalisation power and $R$ is the smoothing scale, and the tidal torque $q_R^2$ is defined as
      \begin{equation}
      \label{eq:alpha2}  
      q_R^2=\frac{1}{2}\left[(\lambda_3-\lambda_2)^2+(\lambda_3-\lambda_1)^2+(\lambda_2-\lambda_1)^2\right]\,,
      \end{equation}
      where $\lambda_1$, $\lambda_2$, and $\lambda_3$ are the eigenvalues of the tidal tensor.
\end{itemize}

We use the galaxy catalogue from \citet{Guo2011} SAM model which is implemented on the subhalo merger tree from the Millennium simulation. The SAM includes physical processes describing galaxy formation and evolution such as gas cooling, star formation, galaxy merger, and feedback from AGN and supernova. Both the halo catalogue and galaxy catalogue are available at the Millennium database\footnote{\url{http://gavo.mpa-garching.mpg.de/Millennium/}}. The galaxy properties we used are:
\begin{itemize}
    \item[(1)] $M_{\rm *}$, the total stellar mass including disk and bulge, in units of $\hmsun$;
    \item[(2)] $\rm SFR$, the star formation rate of galaxy, in units of $\Msun {\rm yr}^{-1}$;
    \item[(3)] $\rm SFR_{\rm bulge}$, the star formation rate in the bulge, in units of $\Msun {\rm yr}^{-1}$;
    \item[(4)] $M_{\rm cold}$, the mass in cold gas disk, in units of $\hmsun$;
    \item[(5)] $M_{\rm hot}$, the mass in in hot gas, in units of $\hmsun$;
    \item[(6)] $R_{\rm cooling}$, the cooling radius, the radius within which the cooling time scale is shorter than the dynamical timescale, in unit of $h^{-1}{\rm Mpc}$;
    \item[(7)] $j_{\rm gas}$, the specific angular momentum of the gas component, in units of $\rm km\, s^{-1} Mpc\, M_\odot^{-1}$;
    \item[(8)] $f_{\rm bulge}$, bulge fraction, defined by the ratio of bulge stellar mass and total stellar mass;
    \item[(9)] $M_{\rm BH}$, the mass of the supermassive black hole at the galaxy center, in units of $\hmsun$.   
\end{itemize}

\subsection{Hydrodynamic simulations}
\label{subsec:hydro}

We also investigate the CCMD in the Illustris hydrodynamic simulation (focused on the Illustris-2) with a box length of 75$h^{-1}$Mpc \citep{Vogelsberger2014,Nelson15}. It adopts the nine-year {\it WMAP} cosmology \citep{Bennett2013} with parameters $\Omega_{\rm m}=0.27$, $\Omega_{\rm b}=0.0456$, $h=0.70$, $n_{\rm s}=0.963$, and $\sigma_8=0.809$. It evolves $910^3$ dark matter particles of mass $3.5\times 10^7\hmsun$ and $910^3$ baryon particles of mass $7\times 10^6\hmsun$ from $z=127$ to $z=0$ with 136 snapshots. At each snapshot, dark matter haloes are identified as FOF groups, and the subhaloes or satellites are identified by the \texttt{SUBFIND} algorithm.

As an improved version of the Illustris simulation, Illustris TNG updates several baryonic processes including AGN feedback, galactic winds, and magnetic fields \citep{Marinacci2018,Naiman2018,Nelson2018,Nelson2019,Pillepich2018,Springel2018}. We focus on the TNG100 simulation, which adopts the Planck cosmology \citep{Planck2016} with $\Omega_{\rm m}=0.31$, $\Omega_{\rm b}=0.0486$, $h=0.677$, $n_{\rm s}=0.97$, and $\sigma_8=0.816$. It has the same box size and initial condition (adjusted for different cosmology) as the original Illustris simulation, allowing for a direct comparison. It contains $1820^3$ dark matter particles of mass $5.1\times 10^6\hmsun$ and $1820^3$ baryon particles of mass $9.4\times 10^5\hmsun$ evolving from $z=127$ to $z=0$ with 100 snapshots. For both Illustris-2 and TNG100 simulations, we only consider the samples of haloes with mass above $\times10^{10.6}\hmsun$.

\subsection{Gaussian mixture model}
\label{subsec:gmm}

In the CCMD framework developed in \citetalias{XuHaojie2018}, which has been applied to model the luminosity/colour-dependent clustering of the SDSS galaxies, the colour-magnitude distribution at fixed halo mass is described by pseudo-red and pseudo-blue components for both central and satellite galaxies. Each of the two central CCMD components is assumed to follow a 2D Gaussian distribution, and the two components are found to be orthogonal to each other in the sense of their different spreads in the luminosity and colour directions. With the best-fit parameters, the CCMD model can reproduce the detailed clustering dependence on luminosity and colour reasonably well. It is then interesting to explore the CCMD components in SAM and hydrodynamic simulations, which would be helpful for understanding their origin and provide insights into galaxy formation and evolution. We mainly use the r-band magnitude and $g-r$ colour without dust extinction in SAM and the simulation output at $z=0$ for investigating the relation between colour and halo or galaxy properties. We also use magnitude and colour including dust extinction for comparisons with observational results.

Since the halo mass is known in SAM and hydrodynamic simulations, it is convenient to examine the existence of the CCMD components directly as a function of halo mass. In this work, we achieve this by decomposing the central colour-magnitude distribution at fixed halo mass bins with Gaussian mixture modelling (GMM). GMM assumes the data consists of two or more Gaussian components of specific dimensions and then infers the means, covariance matrix, and weights of the components by maximizing the likelihood function \citep{Escobar1995}. In practice, the Gaussian parameters are constrained by the expectation maximisation (EM) algorithm which consists of two steps \citep{Dempster1977}. The first step of EM is estimating the expectation of each data point produced by the specific component based on an initial guess of the parameters of the components.  In other words, this roughly classifies the data into different components. Then the second step is tuning the parameters of the components according to the classification in the first step by maximising the expectation of the likelihood. After one iteration, the tuned parameters are treated as input of the next iteration of the two steps until the parameters converge. We apply the GMM model in the \texttt{sklearn} Python package to find the 2D Gaussian components in the colour-magnitude distribution of model galaxies. 

\section{CCMD in SAM}
\label{sec:samCCMD}

We first investigate the CCMD in the \citet{Guo2011} SAM central galaxy sample. Grouping the haloes into mass bins of 0.2 dex from $10^{11.4}\hmsun$ to $10^{14.6}\hmsun$, we show the CCMD components found by the GMM. For the purpose of exploring the origin of the CCMD components, we then focus on the relationship between central galaxy colour and halo or other galaxy properties.

\subsection{GMM results}
\label{subsec:samGMM}
\begin{figure*}
	\centering
	\begin{subfigure}[h]{0.9\textwidth}
	\includegraphics[width=\textwidth]{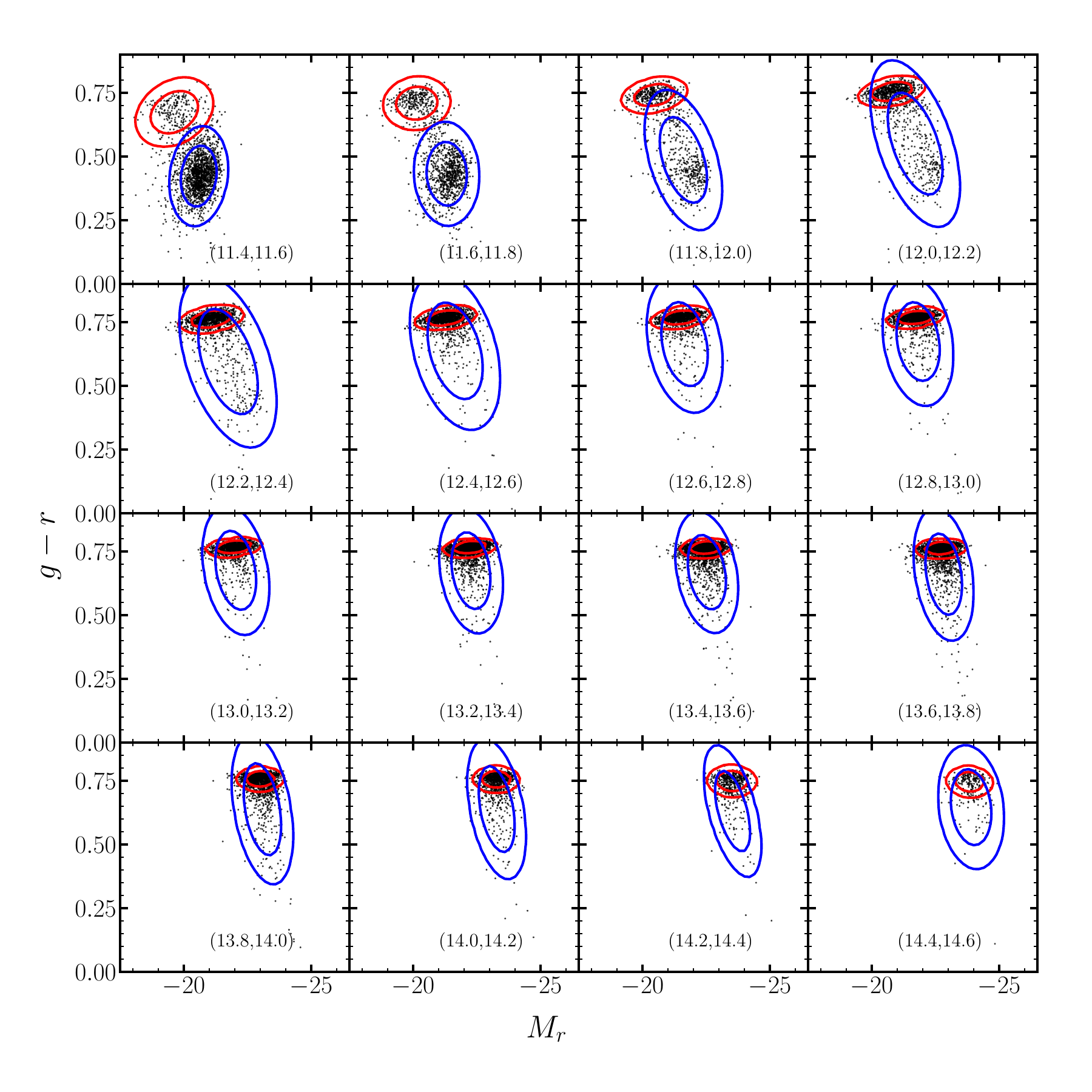}
	\end{subfigure}
\caption{The CCMD components in \citet{Guo2011} SAM central galaxy sample as a function of halo mass. The original colour-magnitude distribution in each mass bin is shown by black dots from randomly selected galaxies. The Gaussian CCMD components found by 2D GMM are shown with red and blue contours for the 1-$\sigma$ and 2-$\sigma$ levels. The logarithmic halo mass range is indicated at the bottom of each panel.}
\label{fig:ccmd_sam}
\end{figure*}

For each halo mass bin, we show the 2D distribution of r-band magnitude $M_{\rm r}$ (on x-axis) and $g-r$ colour (on y-axis) for a randomly selected central sample (black dots) in Fig.~\ref{fig:ccmd_sam}. It can be seen that in each mass bin, the colour-magnitude distribution consists of two populations, a red one and a blue one. In the first two mass bins, the blue population weighs more than the red one, whereas in higher mass bins this is reversed. Since the galaxy colour usually reflects the star formation activity, we note that the halo mass plays a role in galaxy quenching, such that the quenching fraction is lower in low mass haloes and higher in high mass haloes. However, at fixed halo mass bin, the colour difference is likely caused by other halo properties. In this work, we will study the connections between galaxy colour and secondary halo properties or environmental properties at fixed halo mass bins. 

We then apply the 2D GMM to model the colour and magnitude distribution of the centrals in each halo mass bin, and the two Gaussian components found are shown with coloured contours in Fig.~\ref{fig:ccmd_sam}. We find that the colour-magnitude distribution can be decomposed into two distinctive Gaussian components, a red component with a narrow colour range (shown by the red contours) and a blue component with an extended colour range (shown by the blue contours). For halo mass below $10^{12}\hmsun$, the red component can be well separated from the blue component. For halo mass above $10^{12}\hmsun$,  
the red component seems to be well described by the 2D Gaussian, while the 2D Gaussian for the blue component strides across the red Gaussian and overshoots at the red end. However, it is clear that a two-component model provides a reasonable description of the the colour-magnitude distribution. We will use the 2D Gaussian components to inform us about the component characteristics (e.g. the mean colour and magnitude). We also apply the GMM to lower halo mass bins but find that the two components are largely overlapped and indistinguishable. Our results from the SAM are qualitatively in line with the central CCMD components of \citetalias{XuHaojie2018} (their fig.8), constrained from modelling the colour and luminosity dependent SDSS galaxy clustering, such that the colour-magnitude distribution at fixed halo mass can be described as two nearly orthogonal 2D Gaussian components. We note that the red and blue components found here correspond to the pseudo-red and pseudo-blue components in \citetalias{XuHaojie2018}. 

\begin{figure*}
	\centering
	\begin{subfigure}[h]{0.9\textwidth}
	\includegraphics[width=\textwidth]{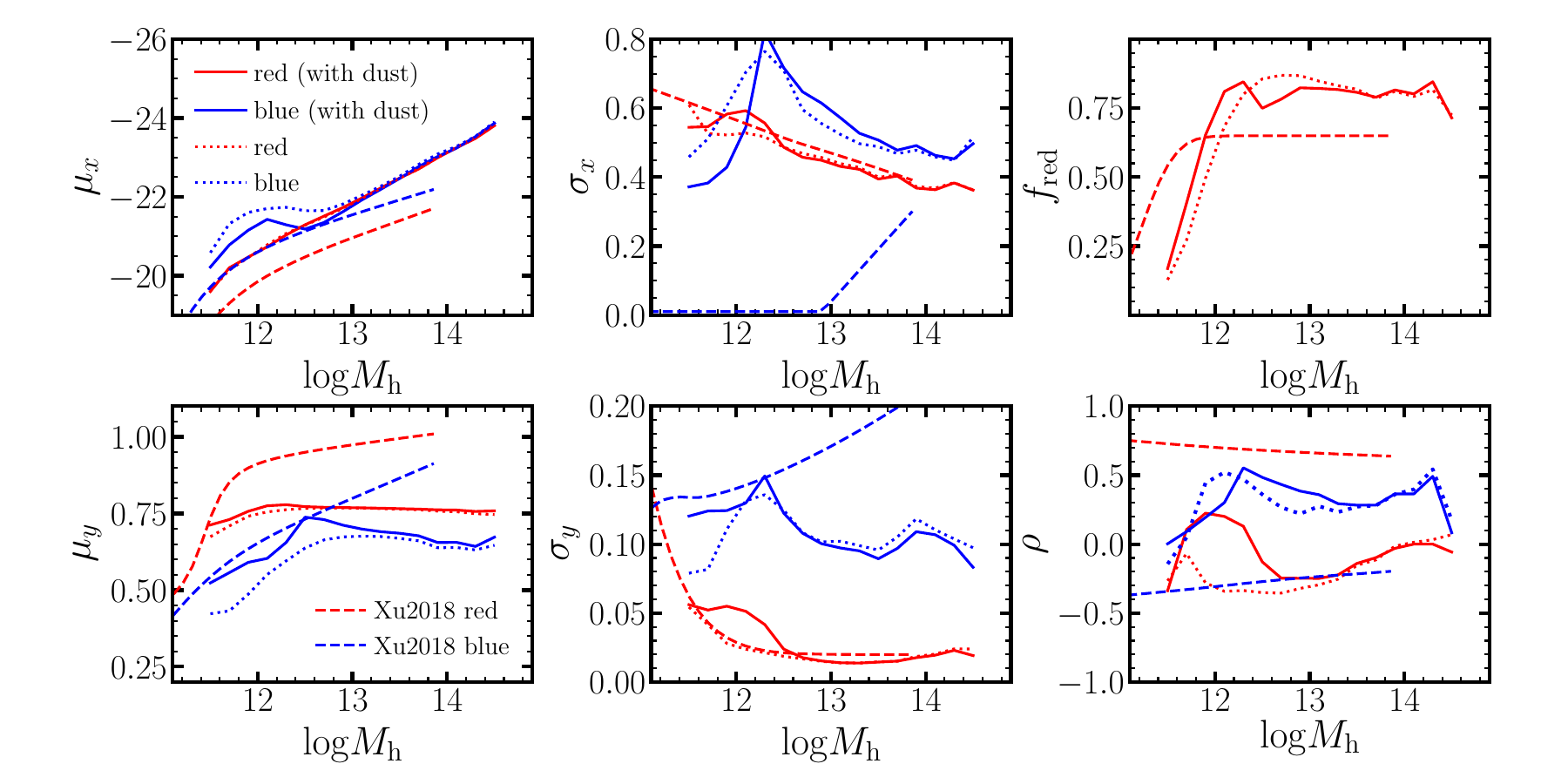}
	\end{subfigure}
\caption{Parameters of the 2D Gaussian components of the CCMD in Fig.~\ref{fig:ccmd_sam} as a function of halo mass. From top left to bottom right, the parameters presented are mean of $M_{\rm r}$ (x-axis in Fig.~\ref{fig:ccmd_sam}), standard deviation of $M_{\rm r}$, fraction of centrals in the red component, mean of $g-r$ colour (y-axis in Fig.~\ref{fig:ccmd_sam}), standard deviation of $g-r$, and the correlation coefficient between $M_{\rm r}$ and $g-r$. The red and blue components are shown by red and blue solid curves respectively. In each panel, solid (dotted) curves show the CCMD parameters from the SAM magnitude and colour in which the dust extinction is included (not included). Dashed curves show the empirically derived results in \citetalias{XuHaojie2018} (based on modelling the luminosity/colour-dependent clustering of the SDSS galaxies). The main qualitative differences between the SAM and the observational result lie in the standard deviation of magnitude (for the blue component) and the correlation coefficients.}
\label{fig:param-mh}
\end{figure*}

From low halo mass to high halo mass, the shapes and positions of the red and blue components gradually change. To investigate this mass dependence in detail, we show the Gaussian parameters of the two components in Fig.~\ref{fig:ccmd_sam} as functions of halo mass in Fig.~\ref{fig:param-mh} by dotted curves. The top-left panel shows the mean of magnitude, with the red (blue) curve for the red (blue) component. The median luminosity, represented by the mean magnitude, increases with halo mass for both the red and blue component. The relation is linear for both red and blue and overlapped in haloes of mass above $10^{13}\hmsun$, indicating a similar mean r-band magnitude for the two components. Below this mass, the mean magnitude of the red component has a nearly linear dependence on ${\rm log}M_{\rm h}$, whereas that of the blue component is nonlinear and has on average more luminous central galaxies. Our Fig.~\ref{fig:param-mh} can be compared to fig.12 of \citetalias{XuHaojie2018}. We include their results here by the dashed curves. For a better comparison, we also show the results using the SAM magnitude and colour in which the dust extinction is included by the solid curves, since these correspond to the magnitude and colour in observation. The results with dust extinction are similar to those without it, and we will focus on the former in Fig.~\ref{fig:param-mh}. \citetalias{XuHaojie2018} also find that both the red and blue components become more luminous with increasing halo mass. However, their blue component appears to be brighter than the red component at all halo masses, and both of their red and blue components are fainter than ours.

The standard deviation of magnitude (shown in the top-middle panel) of the red component decreases linearly with ${\rm log}M_{\rm h}$, consistent with the result in \citetalias{XuHaojie2018}. The standard deviation of magnitude for the blue component increases with ${\rm log}M_{\rm h}$ at low halo mass and decreases with ${\rm log}M_{\rm h}$ at high halo mass. Overall, the red component is narrower in magnitude than the blue component for most of the mass bins, contrary to the trend found in \citetalias{XuHaojie2018}.

The bottom-left panel shows the mean colour of the two CCMD components. The mean colour of the red component keeps approximately constant along the whole halo mass range, while the mean colour of the blue component increases with halo mass below $\log M_{\rm h}<12.5$ then slowly decreases above this mass. In \citetalias{XuHaojie2018}, the increase in mean colour with halo mass can be seen more clearly for both the two components in the whole mass range. 
The bottom-middle panel shows the standard deviations of colour of the two components as functions of halo mass. In the whole mass range, the standard deviation of colour of the blue component is higher than that of the red component, indicating a wide colour range of the blue component and a narrow colour range of the red component. This is consistent with that found in \citetalias{XuHaojie2018} (above $\log M_{\rm h}\sim 11.5$). The standard deviation of colours of the red component in SAM is very similar to that in \citetalias{XuHaojie2018}, while that of the blue component is lower than that in \citetalias{XuHaojie2018}, especially in massive haloes. Compared to the mean and standard deviation of magnitude,  which show strong dependences on halo mass in the whole mass range, the mean and standard deviation of colour depend weakly on halo mass above ${\rm log}M_{\rm h}=13$. 

The fraction of centrals in the red component in each mass bin is provided in the top-right panel. The red fraction roughly has a linear dependence on ${\rm log}M_{\rm h}$ below ${\rm log}M_{\rm h}<12.5$, increasing from 0.2 to 0.8, and then becomes approximately constant at $\sim 0.8$ above this mass. This is very similar to that in \citetalias{XuHaojie2018}, except that the high-mass plateau of their red fraction is at $\sim0.6$.

The bottom-right panel shows the correlation coefficient between colour and magnitude for the blue and red components, where zero indicates no correlation and positive (negative) value indicates positive (negative) correlation. In general, magnitude and colour are negatively (positively) correlated for the red (blue) component with negative (positive) correlation coefficient. More luminous centrals with more negative magnitude tend to be redder (higher colour value) in the red component, whereas dimmer centrals with less negative magnitude tend to be redder in the blue component. The absolute values of the correlation coefficients are small, suggesting a weak correlation between magnitude and colour. Interestingly, \citetalias{XuHaojie2018} find an opposite trend inferred from observation that magnitude and colour are positively (negatively) correlated for the red (blue) component, with  stronger correlations than what we find here.
 
We also perform the same analysis with both narrower (0.1 dex) and wider (0.4 dex) halo mass bins and find that all the results of red and blue components shown in Fig.~\ref{fig:ccmd_sam} and Fig.~\ref{fig:param-mh} remain the same. The overall similar trends in the central galaxy CCMD seen in the SAM and in \citetalias{XuHaojie2018} are encouraging. The differences between our results and \citetalias{XuHaojie2018} are likely caused by the details of galaxy formation mechanisms employed in the SAM compared to the real universe. 

While it is difficult to link the CCMD components to halo properties or galaxy formation processes from observation,  given that the halo properties and some of the galaxy properties can not be directly observed, with the SAM we can investigate such connections, as we do in the following subsections.

\subsection{Correlation between CCMD or galaxy colour and halo properties}
\label{subsec:colour_and_halo}

\begin{figure*}
	\centering
	\begin{subfigure}[h]{0.9\textwidth}
	\includegraphics[width=\textwidth]{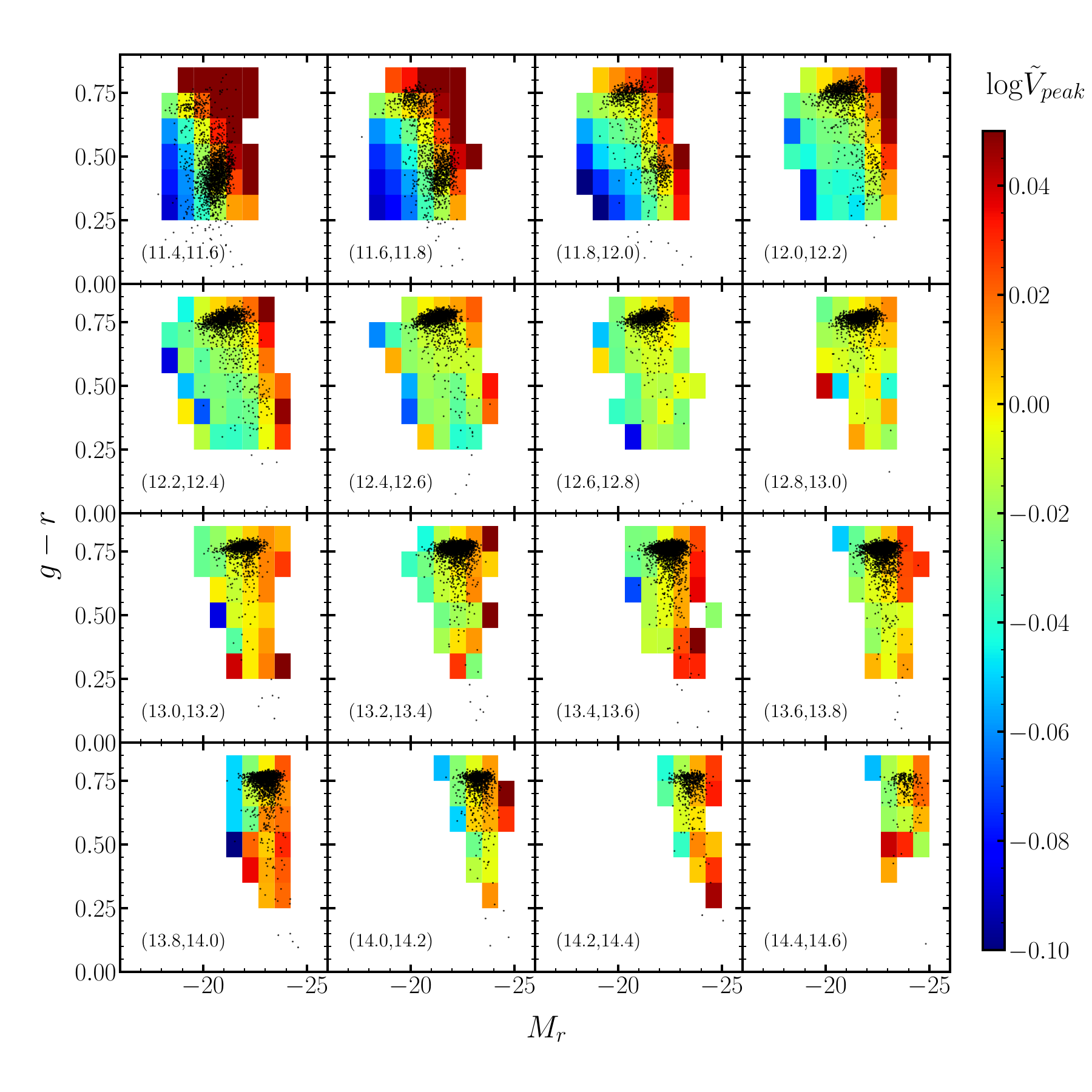}
	\end{subfigure}
	\hfill
\caption{$V_{\rm peak}$ variation as a function of central galaxy magnitude and colour. Same as Fig.~\ref{fig:ccmd_sam}, each panel shows one halo mass bin, which is further split into colour-magnitude bins. Colour scale represents normalised quantity $\tilde{V}_{\rm peak}$, defined as the ratio of $V_{\rm peak}$ in a colour-magnitude bin and its mean in the halo mass bin. Red (blue) colour indicates relative high (low) value of $\tilde{V}_{\rm peak}$ in the halo mass bin. The black dots show randomly selected samples of central galaxies.}
\label{fig:ccmd-con}
\end{figure*}

We now investigate the relationship between the central CCMD components and halo properties in the SAM sample. We show the variation of $V_{\rm peak}$ in the conditional colour-magnitude plane in Fig.~\ref{fig:ccmd-con}. The halo mass bins are the same as those in Fig.~\ref{fig:ccmd_sam}. In each halo mass bin, we further split the centrals in colour-magnitude bins and measure the mean of a normalised quantity $\tilde{V}_{\rm peak}$, which is defined as the ratio of $V_{\rm peak}$ in a colour-magnitude bin and its mean in the halo mass bin. The advantage of using the normalised $V_{\rm peak}$ is that it spans similar range for different halo mass bins. The colour coding in each panel shows the mean $\tilde{V}_{\rm peak}$ across the colour-magnitude bins. For haloes of ${\rm log}M_{\rm h}<13$, $V_{\rm peak}$ varies along the diagonal direction in colour-magnitude plane in a way that brighter redder centrals tend to live in haloes with higher $V_{\rm peak}$, and fainter bluer centrals tend to live in haloes with lower $V_{\rm peak}$. This is understandable, since at fixed halo mass, haloes with higher $V_{\rm peak}$ usually form earlier, and the time for centrals to grow and evolve is longer. This indicates that for haloes in this mass range, both central magnitude and colour depend on halo assembly history. For haloes above ${\rm log}M_{\rm h}>13.0$, $V_{\rm peak}$ varies essentially along the magnitude direction, in a way that brighter centrals are hosted by haloes with higher $V_{\rm peak}$, and the variation with colour is rather weak. In other words, compared to magnitude, central galaxy colour in massive haloes is almost independent of halo assembly history. Same as Fig.~\ref{fig:ccmd_sam} and Fig.~\ref{fig:param-mh}, the results in Fig.~\ref{fig:ccmd-con} are not sensitive to the width of halo mass bins. Similar variation trends are found for other halo assembly properties such as $a_{\rm 0.5}$ and concentration.

Fig.~\ref{fig:ccmd-con} shows different dependences of central colour and magnitude on $V_{\rm peak}$ for haloes of different masses. However, compared to Fig.~\ref{fig:ccmd_sam}, this is inadequate to explain the two CCMD components, since the presence of the CCMD components is independent of the variation of $V_{\rm peak}$. As seen in Fig.~\ref{fig:ccmd_sam}, the major difference between red and blue components at fixed halo mass is the $g-r$ colour, with a similar range of $M_{\rm r}$ magnitude. Exploring the origin of the CCMD components in terms of galaxy-halo connection is thus approximately equivalent to studying the relation between galaxy colour and halo properties.

\begin{figure*}
	\centering
	\begin{subfigure}[h]{0.9\textwidth}
	\includegraphics[width=\textwidth]{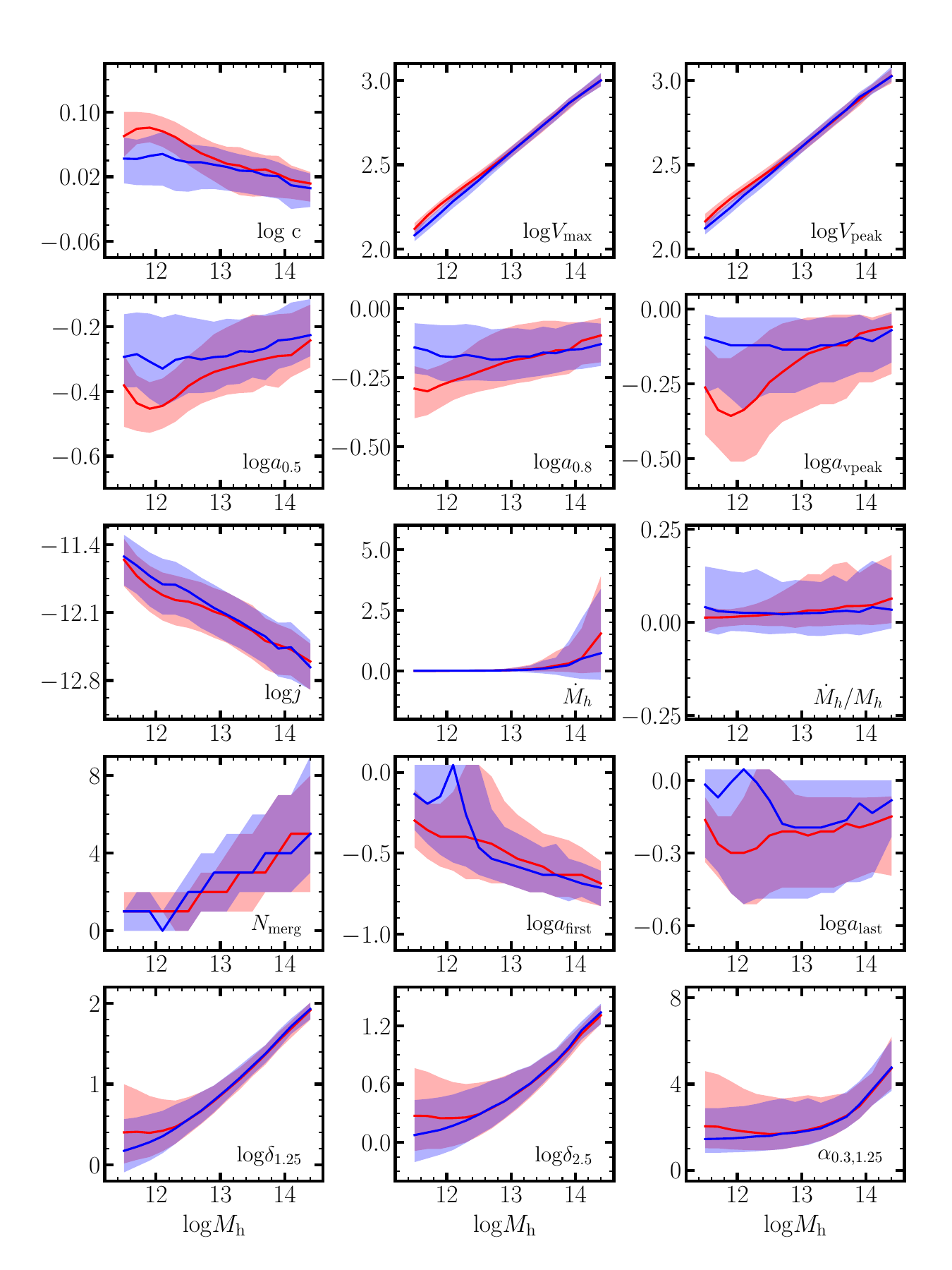}
	\end{subfigure}
	\hfill
\caption{Secondary halo properties defined in section~\ref{subsec:guo2011} as a function of halo mass for extreme red and blue central galaxies. The extreme red/blue galaxies are the lower/upper 10\% percentiles of colour in each mass bin. The halo properties are labelled at the bottom right corner in each panel, and the units are shown in section~\ref{subsec:guo2011}. The first four rows show halo internal properties, and the last row shows environmental properties. The red/blue solid curves show the median of the specific halo property of the extreme red/blue population in each halo mass bin, and the red/blue bands indicate the 16\%-84\% percentiles.}
\label{fig:colorsep}
\end{figure*}

To clearly see the difference in halo properties for galaxies of different colours, In each mass bin, we select extreme red and extreme blue populations of central galaxies as the lower/upper 10\% percentiles of colour. We combine the last two mass bins for a reasonable number in the two populations. Fig.~\ref{fig:colorsep} shows the halo properties as a function of halo mass, separated by these two populations. The halo properties considered are listed in Section.~\ref{subsec:guo2011}, and are labelled at the bottom right corner in each panel. The solid red (blue) curve presents the median of the halo property in each halo mass bin, and the red (blue) band around it encloses the 16\%--84\% percentiles of the halo property. 

In the whole halo mass range, the haloes hosting extreme red centrals have higher concentrations compared to those hosting extreme blue centrals, and the trend is more pronounced at low mass. Similarly, in low mass haloes ($M_{\rm h}\lesssim 10^{13}\hmsun$), the haloes hosting extreme red centrals have higher $\vmax$ and $\vpeak$ compared to those hosting extreme blue centrals. The haloes with higher concentration, $\vmax$, or $\vpeak$ are typically older haloes (see e.g. \citealt{Xu2018}), which have lower $\ahalf$ (the cosmic scale factor when a halo reaches half of its final mass; see the left panel on the second row). That is, in the SAM, at a fixed halo mass, extreme red centrals tend to reside in haloes that form earlier. Similar to $\ahalf$, $a_{\rm 0.8}$ and $a_{\rm vpeak}$ also characterise the halo formation history. They are the cosmic scale factors when the mass of a halo reaches 0.8 times of its final mass and its maximum circular velocity reaches $\vpeak$, respectively. Their correlations with galaxy colour are consistent with that of $\ahalf$. 

There is a weak trend that the extreme blue centrals reside in haloes with higher specific angular momentum, but the difference in the specific angular momentum between the extreme red and blue populations is small, given the overlap in the distributions. For either the mass accretion rate or the specific mass accretion rate of haloes, it is difficult to distinguish between the two populations of extreme colours. This is contrary to the naive expectation that a higher mass accretion rate would enhance star formation. The reason can lie in the fact that many other processes may also affect star formation from accreted gas, such as AGN activities that we will explore in the following section. 

The fourth row of Fig.~\ref{fig:colorsep} displays the relation between the two populations of extreme colour and halo major merger. The total number of major mergers of haloes hosting extreme blue centrals is slightly higher than that of haloes hosting extreme red centrals. At the low halo mass end, the haloes hosting extreme blue centrals experience the first and the last major merger later than those hosting extreme red centrals. However, the distribution of each of the major merger properties is wider than the difference, and it appears hard to tell apart the two populations of galaxies solely from the major merger properties.

In addition to the halo internal properties, environmental properties are also believed to affect galaxy formation and evolution. The environment can refer to both halo-scale environment and large-scale environment outside the halo. Using an SDSS group catalogue, \citet{Woo2013} illustrate that satellite galaxy quenching fraction can depend on halo-scale environment such as halo mass and distance from the halo center. They find that central quenching fraction also depends on halo mass and depends weakly on large-scale environment. Here we focus on large-scale environment outside haloes and explore the relationship between environment and central galaxy colour at fixed halo mass. Results are present in the bottom panels of Fig.~\ref{fig:colorsep}. The environment properties we considered here are $\delta_{\rm 1.25}$ and $\delta_{\rm 2.5}$ (matter density smoothed on scales of 1.25 and 2.5 $\hmpc$, respectively), as well as the tidal anisotropy $\alpha_{\rm 0.3,1.25}$ measured with the $\delta_{\rm 1.25}$ field. Since the density field is smoothed with a Gaussian filter, the mass enclosed is approximately within the radius of $\sqrt{5}\times1.25$ or $\sqrt{5}\times 2.5 \hmpc$, larger than the typical one-halo scale. The effects of environment (although weak) are mainly seen for low mass haloes, where extreme red (blue) centrals live in denser (lower density) regions. On the other hand, the environmental effect is negligible for haloes of ${\rm log}M_{\rm h}>12$, which is consistent with the result from \citet{Woo2013}. 

To conclude, in the SAM catalogue we use, some of the internal properties (e.g. formation time and concentration) of host haloes show differences for extreme red and blue centrals. These differences are mainly seen for haloes with masses lower than $10^{12.5}\hmsun$. This is usually considered as galaxy assembly bias, i.e. galaxy property dependence on secondary halo properties \citep{Zehavi2018,Xu2020}. The differences for high mass haloes are small if any. Halo properties investigated here do not seem sufficient to explain the central galaxy colour, since baryonic processes can also contribute to the colour evolution. We will explore this in the following subsection by investigating the connection between central galaxy colour and other galaxy properties. 

\subsection{Correlation between galaxy colour and galaxy properties}
\label{subsec:colour_and_gal}

\begin{figure*}
	\centering
	\begin{subfigure}[h]{0.9\textwidth}
	\includegraphics[width=\textwidth]{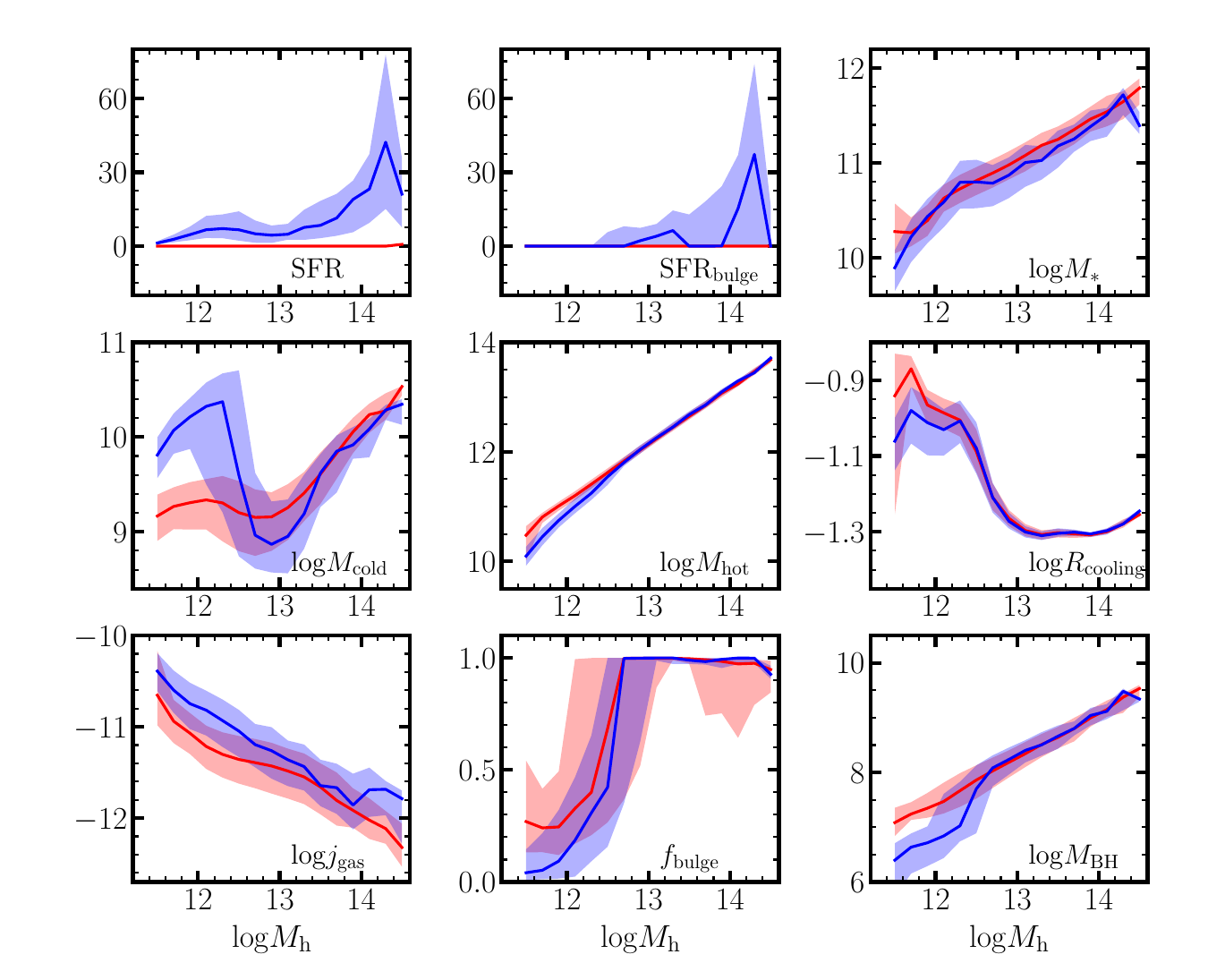}
	\end{subfigure}
\caption{Similar to Fig.~\ref{fig:colorsep}, but for galaxy properties instead of halo properties. Units can be found in section~\ref{subsec:guo2011}.
}
\label{fig:colorsep-galprop}
\end{figure*}

We perform an analysis similar to that in section~\ref{subsec:colour_and_halo}, replacing the halo properties with galaxy properties. Fig.~\ref{fig:colorsep-galprop} shows the galaxy properties as functions of host halo mass separated by extreme red and extreme blue centrals.
We first show the galaxy SFR in the top-left panel. Not surprisingly, the extreme red centrals selected by our colour cut are all quenched with nearly zero SFR, and the extreme blue centrals have non-zero SFR. The panel to the right shows the SFR in the bulge instead of the total SFR. Different than the total SFR, the bulge SFR of extreme blue centrals decreases to zero in some halo mass bins, which reveals that the galaxy colour might be more directly correlated with SFR in the disk for those haloes. The relation between colour and stellar mass is shown in the top-right panel. At fixed halo mass, the extreme red centrals are slightly more massive than the extreme blue centrals. We note that the correlation between colour and stellar mass is weak, especially at low halo mass. This is in agreement with with \citet{Xu2020}, who also find that in the Illustris simulation the central colour is nearly independent of stellar mass (i.e. tight correlation between stellar mass and halo mass) but correlates more tightly with specific star formation rate.

The differences in the amount of cold/hot gas and in the cooling radius between extreme red and extreme blue centrals are presented in the second row of Fig.~\ref{fig:colorsep-galprop}. In low-mass haloes, there are a larger amount of cold gas in extreme blue centrals than in extreme red centrals, and the trend is opposite for the amount of hot gas. This is reasonable, since stars form from cold gas. However, in high-mass haloes the amount of either cold or hot gas becomes indistinguishable between the two populations with extreme colours. The cooling radius is defined by the radius at which the cooling time scale is shorter than the dynamical time scale. In low-mass haloes, the cooling radii of extreme red centrals are larger than those of extreme blue centrals. The cooling radii of the two extreme colour populations are also similar in high-mass haloes and approximately approach a constant above ${\rm log} M_{\rm h}=13$.

The spin of galaxy could also relate to galaxy colour, since spiral galaxies with higher spin are usually young and star forming and elliptical galaxies are typically old with quenched star formation. In the bottom-left panel of Fig.~\ref{fig:colorsep-galprop}, we show the relation between central colour and specific angular momentum of gas. In the whole halo mass range, gas in extreme blue centrals has higher specific angular momentum than that in extreme red centrals. The connection between gas and halo angular momentum can be complex. The initial angular momentum of the host halo results from tidal torque from the environment and subsequent mergers may change the angular momentum direction in some random way. Tidal disruption and galaxy merger may further randomise the angular momentum direction of gas in galaxies. If gas angular momentum plays a non-negligible role in determining galaxy colour at $z=0$, it would be difficult to connect galaxy colour to host halo properties. 

We then show the relation between colour and bulge fraction in the bottom-middle panel. We define the bulge fraction as the ratio of bulge stellar mass to total stellar mass in the galaxy, and it increases with halo mass. For haloes below ${\rm log} M_{\rm h}=12.6$, extreme red centrals have a slightly larger bulge fraction than extreme blue centrals. Above ${\rm log} M_{\rm h}=12.6$, bulge fraction reaches unity for both the two populations.

Finally, we show the relation between galaxy colour and the mass of the central supermassive black hole. Black hole mass is related to the strength of AGN feedback that could impact the star formation activity in galaxy. For haloes with ${\rm log} M_{\rm h}<12.5$, the central black hole mass of extreme red galaxies at fixed halo mass are higher than those of extreme blue galaxies. It is possible that a more massive black hole could have a higher accretion rate that leads to a higher radio-mode feedback applied in the  \citet{Guo2011} SAM and thus slows down the star formation activity in the galaxy, resulting in an extreme red colour. These active galactic nuclei (AGN) processes are efficient at ${\rm log} M_{\rm BH}>7$, which corresponds to  ${\rm log} M_{\rm h}>11$. However, this effect is not seen for high-mass haloes with ${\rm log} M_{\rm h}>12.5$. Black hole mass in extreme red and blue galaxies above this halo mass are very similar, indicating that black hole or AGN feedback may not be a dominant factor of galaxy colour at high halo mass. Our results are in agreement with the recent work by \citet{Cui2021}. With hydrodynamic simulations, they focus on the regime of ${\rm log} M_{\rm h}<12.5$ and find that the colour bimodality is related to cold gas content, which relies on AGN feedback. 

Similar to the relations between colour and halo properties in section~\ref{subsec:colour_and_halo}, we find that the differences in galaxy properties for extreme red and blue are also mainly seen for haloes with $M_{\rm h}<10^{12.5}\hmsun$. Above this mass the difference is small, except for the specific angular momentum of gas, which shows partially overlapped separation in the whole mass range. For high-mass haloes, it is difficult to fully explain the central colour with the halo properties and galaxy properties we focus on in this work. It is possible that the connections between them are complicated in high dimensions and may not be well separated in the one-dimensional (1D) projection of halo and galaxy properties for different colour populations. As the next step, we will investigate this high dimensional connection with 
an alternative method and explore the relative importance of halo and galaxy properties for determining central galaxy colour.

\section{Predicting central galaxy colour with machine learning}
\label{sec:RF}

From section~\ref{subsec:colour_and_halo} and section~\ref{subsec:colour_and_gal}, we show that central galaxy colour could depend on multiple halo and galaxy properties in a complicated way. To study the multi-variate dependence of central galaxy colour on halo and galaxy properties, we resort to a machine learning technique. In general, machine learning methods have advantages in learning multivariate non-linear dependence and have been extensively implemented for astronomical and cosmological studies. In addition to learning the relation between input properties (or features) and the output (or target label), some machine learning algorithms, for example, the random forest (RF; \citealt{Breiman2001}), also provide the relative importance of the input features for learning the output. In this section, we aim at predicting central galaxy colour with RF using halo and galaxy properties as input. This will also provide us intuitions on which properties are important for determining the central galaxy colour.

RF learns the final output based on a large number of decision trees \citep{Breiman1984}, which is also a machine learning algorithm. As a data driven method, decision tree is built from a root that contains all training samples. When the tree grows top-down, branches are created from the root by splitting the training sample into sub-samples according to selection criteria of one input feature. The splitting feature is chosen by minimising a specific cost function, and the sub-samples are then placed in daughter nodes. The cost function is usually the weighted sum of the Gini impurity from the daughter nodes for classification or the mean squared error for regression. For each daughter node, the splitting is implemented again (based on a different input feature) to separate it into next level sub-samples. The iteration continues and the tree grows down until the number of samples in a node reaches a minimum (i.e. a leaf) or the maximum depth of the tree is achieved. The input features and target label for the training samples are not modified through the decision tree. They are only split into sub-samples flowing down the tree branches. To the end, the target label of each leaf is set by the target label of the majority in the leaf for classification tree or the mean for regression tree. Once the decision tree is established, new test data without target label can go through the tree and obtain predicted label at the leaf. 

RF contains a large number of decision trees, and the output of the RF is an average of the results from all decision trees. We adopt the machine learning package \texttt{sklearn} in Python to perform the RF prediction of central galaxy colour, with input features being halo and galaxy properties. We randomly select 80\% of the galaxies in the central sample as training data and the other 20\% as test data. To obtain the best-fit hyper-parameters such as the maximum depth of the tree, the minimum number of samples in leaf, and the number of decision trees in RF, we perform a grid search over these parameters. The best-fit hyper-parameters are those leading to the best prediction performance estimated by the $R^2$ score \citep{Helland1987,Cox1989}. To avoid overfitting, we split the training sample into three sub-samples. Two of them are used to train the RF model, and the remaining one is used for cross-validation. We train two RF models to predict central galaxy colour, one including only halo properties as input and the other one including both halo and galaxy properties.

\subsection{Colour prediction based on halo and galaxy properties}
\label{subsec:RF_color}

\begin{figure*}
	\centering
	\begin{subfigure}[h]{0.9\textwidth}
	\includegraphics[width=\textwidth]{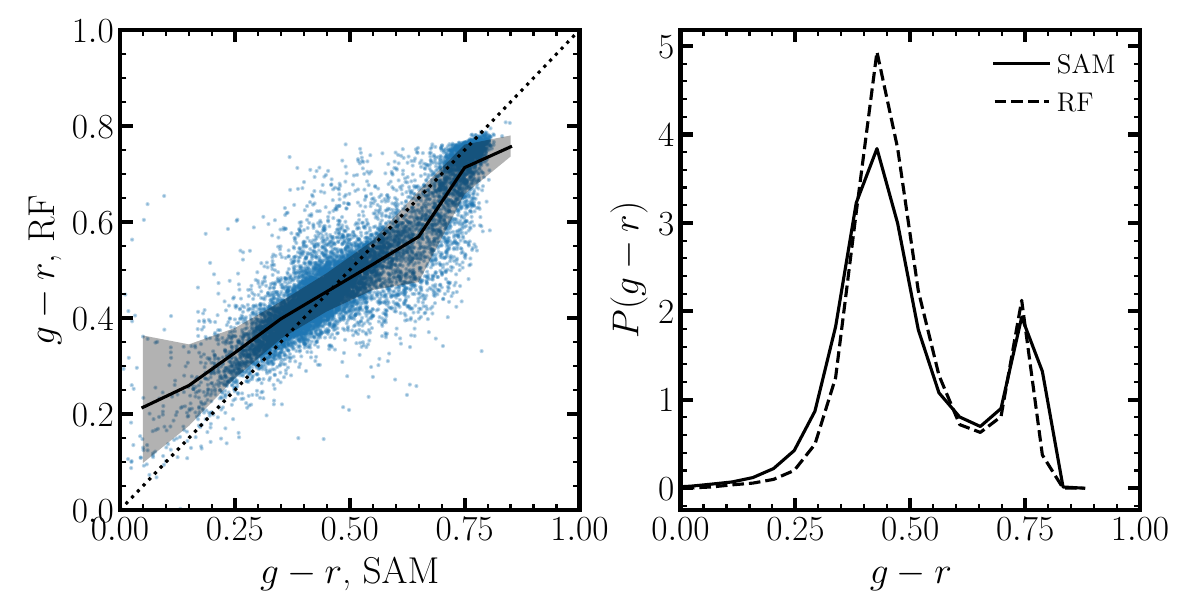}
	\end{subfigure}
	\hfill
\caption{Comparison between central galaxy colour from SAM and RF prediction based on halo properties. Left: individual comparison shown by blue dots and mean (scatter) shown by black solid (grey band). Right: distribution of central galaxy colour, where the solid (dashed) curve shows the SAM (predicted) colour.
}
\label{fig:compare-color-haloRF}
\end{figure*}

We present the RF prediction result trained by the SAM sample for central galaxy colour in Fig.~\ref{fig:compare-color-haloRF}, with the input features being the halo internal and environmental properties shown in Fig.~\ref{fig:colorsep}. In the left panel, comparison to the original SAM colour is shown for individual haloes by the blue dots. The majority of the data points are distributed along the equality line (black dotted), but clear scatters can be seen. We then divide the haloes into bins of SAM colour and measure the mean (black solid) and scatter (grey band) of the RF predicted colour to show the discrepancy quantitatively. At the blue end of  SAM colour, the RF tends to predict the colour to be redder, and the trend reverses for red galaxies. The $R^2$ score for the prediction is $\sim$0.7, suggesting that 70\% of the variance in the original colour can be explained by the RF prediction (a unity $R^2$ score indicates a perfect prediction). 
Comparing the overall distribution of the RF predicted colour to that in the SAM in the right panel shows that the prediction produces fewer extreme blue centrals at $g-r\lesssim 0.4$ and more centrals around the peak of $g-r\sim0.4$. It also predicts fewer extreme red colour at $g-r\gtrsim 0.8$. The statistical properties of the SAM colour are not reproduced very accurately. It is possible that using only the halo properties as input is inadequate for the RF prediction.   

\begin{figure*}
	\centering
	\begin{subfigure}[h]{0.9\textwidth}
	\includegraphics[width=\textwidth]{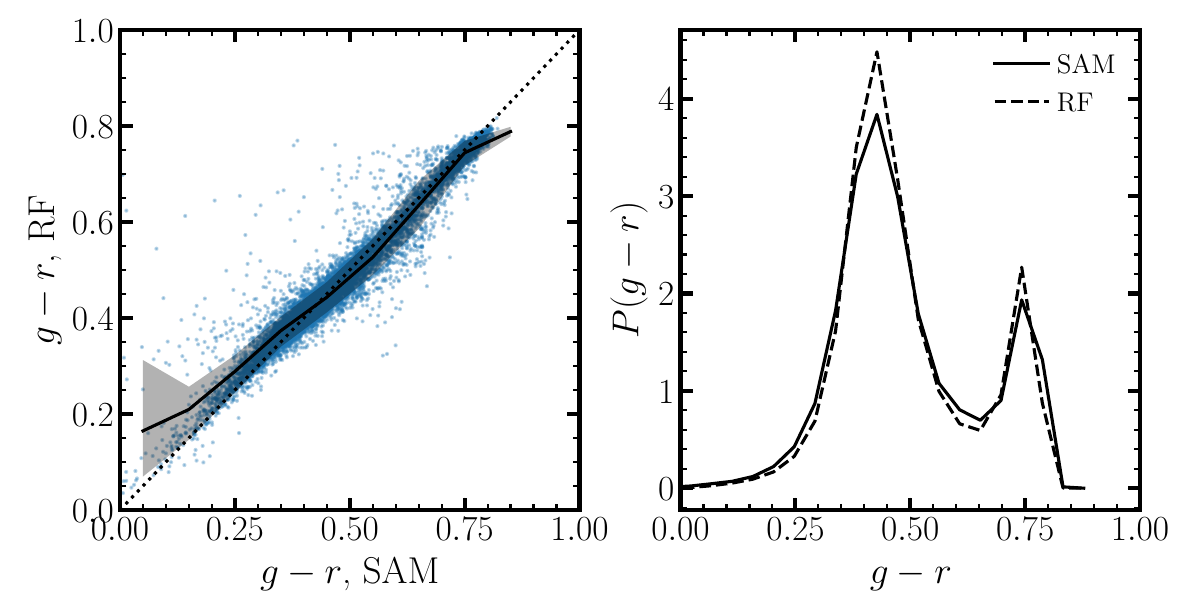}
	\end{subfigure}
	\hfill
\caption{Similar to Fig.~\ref{fig:compare-color-haloRF}, but for the comparison between SAM colour and RF prediction based on both halo and galaxy properties.
}
\label{fig:compare-color-halogalRF}
\end{figure*}

We then add galaxy properties shown in Fig.~\ref{fig:colorsep-galprop} (except for the SFR quantities) to the input of the RF model. After re-tuning the hyper-parameters, we train the new RF model to predict the central colour and show the results in Fig.~\ref{fig:compare-color-halogalRF}. The prediction performance of the new RF is largely improved with an $R^2$ score $>0.9$. The scatter and extreme discrepancy are both reduced for the individual comparison (blue dots) between the RF prediction and SAM colour compared to the left panel of Fig.~\ref{fig:compare-color-haloRF}. In terms of the overall distribution, the level of discrepancy between the RF prediction and the SAM is also largely reduced. This indicates again that the halo properties alone are inadequate to recover the central colour reasonably well, and the galaxy properties (or baryonic processes) have a large contribution to the central colour.

\begin{figure*}
	\centering
	\begin{subfigure}[h]{0.95\textwidth}
	\includegraphics[width=\textwidth]{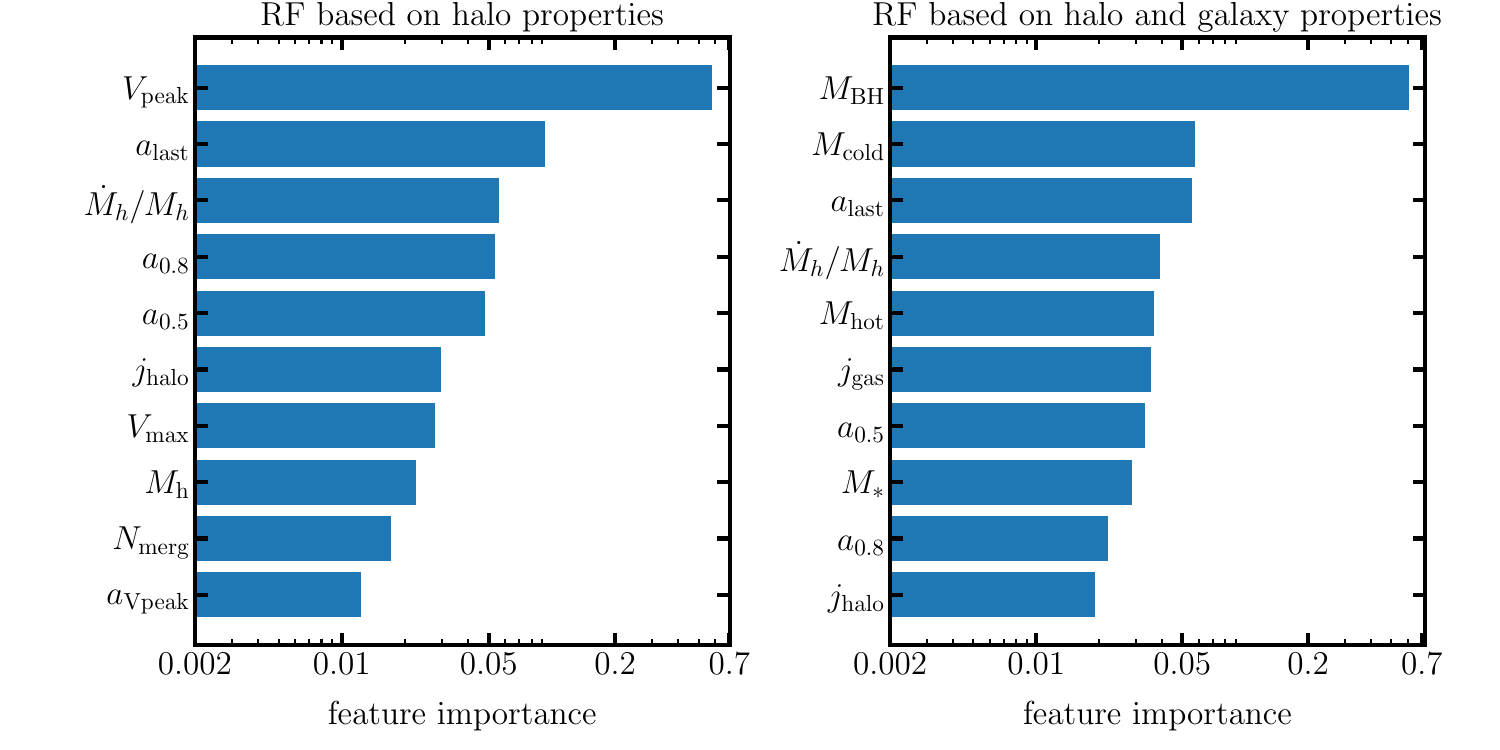}
	\end{subfigure}
	\hfill
\caption{Feature importance of halo or galaxy properties from the RF models for predicting central galaxy colour. Left and right panels show the RF model based on halo properties only and on both halo and galaxy properties, respectively. Only the top ten important properties are shown, which are ordered by their relative importance.
}
\label{fig:feature-importance}
\end{figure*}

In addition to the predicted galaxy colour, the RF algorithm also provides an estimation of feature importance. This estimation is based on the contribution of the input features to the construction of decision trees. We present the feature importance for the halo properties-only RF and the RF including both galaxy and halo properties in Fig.~\ref{fig:feature-importance}. For both RF models, we select the top ten important features and order them by the relative importance. For halo properties-only RF, the most important property is $\vpeak$. This is not surprising since $\vpeak$ is not only a halo mass indicator but also related to the assembly history of the halo. $\vmax$ is also an input feature that is highly correlated with $\vpeak$. However, its relative importance is much lower than that of $\vpeak$, ranking as the seventh feature. This is likely caused by the design of the feature importance algorithm instead of a true difference in importance between the two properties. For example, if we remove $\vpeak$ from the input features, $\vmax$ will be the most important feature with a similar relative importance value ($\sim$0.7). So the true importance of $\vpeak$ and $\vmax$ are likely to be similar. Since the training uses haloes in the whole mass range, we include $M_{\rm h}$ as one of the input features, and it is ranked as the eighth important property, lower than $\vpeak$ and $\vmax$. This is likely due to both the tight correlation with $\vpeak$ and the lack of assembly information contained in $M_{\rm h}$. Halo formation history properties such as the scale factor $a_{\rm last}$ when the halo experiences the last major merger, the scale factor $a_{\rm 0.8}$ when the halo reaches 0.8 of final mass, and the scale factor $a_{\rm 0.5}$ when the halo reaches 0.5 of final mass, rank from the second to the fifth, indicating that the central colour is correlated with halo formation history. This is consistent with Fig.~\ref{fig:colorsep} that the host halo of extreme red and blue centrals are more distinguishable in terms of these formation history properties. Ranking as the third important feature, the specific mass accretion rate also has some predictive power in the central galaxy colour.  

We note that all the top 10 properties listed here are halo internal properties, and none of the environmental properties is shown on the list. Therefore, we consider the large-scale environment as less important for determining the central galaxy colour. With the same SAM model, \citet{Zehavi2018} show that the dependence of central stellar mass on the environment is much weaker than that on halo formation time. Using similar RF prediction, \citet{Xu2021b} find that environment properties are less important than halo internal properties for predicting central galaxy occupation above a specific stellar-mass threshold. Together with these results, we infer that the formation and evolution of central galaxies depend more on halo formation history than the large-scale environment.  

The right panel of Fig.~\ref{fig:feature-importance} shows the rank of feature importance when both halo and galaxy properties are included. The mass of the central supermassive black hole, which can be considered as an indication of AGN feedback strength, is now ranked as the most important property. In the \citet{Guo2011} SAM model, the AGN feedback strength (e.g. the rate of energy input to the atmosphere) is defined to depend on the hot gas accretion rate of the black hole, which is proportional to the black hole mass. So gas cooling in centrals with more massive black holes would be affected more strongly by the AGN feedback energy, which would be more effective to slow down the star formation activity. This is consistent with the trend found in Fig.~\ref{fig:colorsep-galprop}, where the extreme red centrals have higher black hole mass than the extreme blue centrals in low mass haloes. Our result is in agreement with \citet{Cui2021}, who find that the central colour bimodality relies on Jet-mode AGN feedback and X-ray mode AGN feedback in the SIMBA hydrodynamic simulation \citep{Dave2019}. As the second important property, the mass of cold gas is also a major determining factor for star formation. The third and the fourth important properties are halo formation history property and accretion rate property. We note that $V_{\rm peak}$ or $V_{\rm max}$ are not among the top 10 important properties when including galaxy properties. A possible reason is that their information may be included by galaxy properties such as black hole mass, which correlates tightly with them. The Pearson correlation coefficient between black hole mass and $V_{\rm peak}$/$V_{\rm max}$/$M_{\rm h}$ is 0.7/0.7/0.8. Overall, the galaxy properties are shown to be more important than the halo properties for predicting the central colour. 

We note that all the feature importance results could be dominated by low-mass haloes, since their abundance is much higher than high-mass haloes, leading to more contribution to the construction of the RF model. We also perform a test of training separate RF models for ${\rm log} M_{\rm h}<12.5$ and ${\rm log} M_{\rm h} \geq 12.5$. The RF model for the latter has a low $R^2$ score, which means it barely predicts the central galaxy colour. The feature importance inferred from such a model is therefore not reliable. This is consistent with the results in section ~\ref{subsec:colour_and_halo} and section ~\ref{subsec:colour_and_gal} that the central colour correlates weakly with halo and galaxy properties for high halo masses. With the halo and galaxy properties we focused on here, we are not able to find conclusively the most significant ones for determining the central galaxy colour in high-mass haloes.

\subsection{CCMD prediction based on halo properties only}
\label{subsec:RF_colormag}

\begin{figure*}
	\centering
	\begin{subfigure}[h]{0.9\textwidth}
	\includegraphics[width=\textwidth]{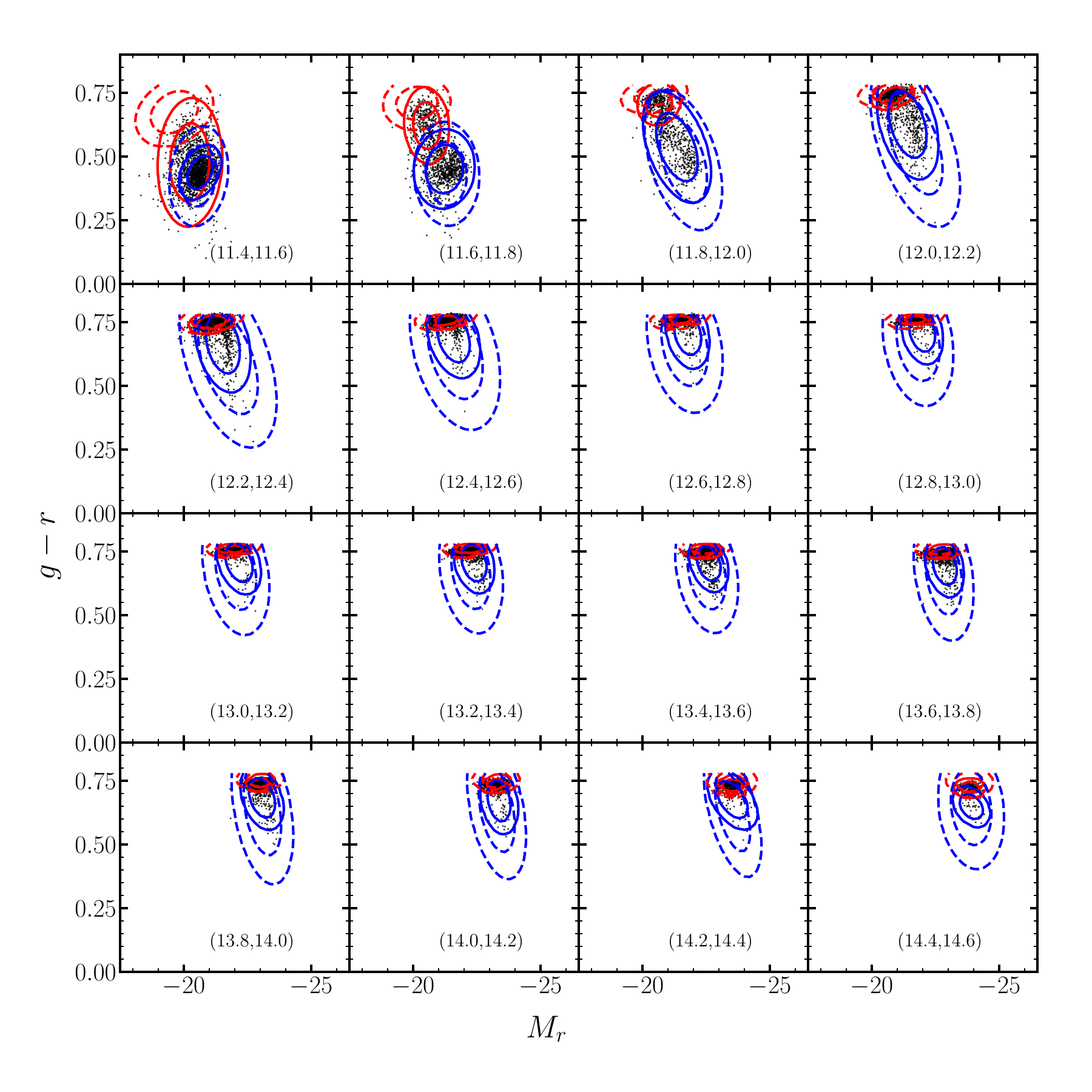}
	\end{subfigure}
	\hfill
\caption{CCMD in RF predicted $M_{r}$ and $g-r$ based on halo properties (solid) compared to that in the SAM (dashed). Black dots show randomly selected RF predictions.}
\label{fig:ccmd_pred}
\end{figure*}

In section~\ref{subsec:RF_color} we explore the relation between central galaxy colour and halo or galaxy properties based on the RF model prediction. Since the CCMD is a 2D distribution of colour and magnitude, we also perform a similar RF model for the r-band magnitude $M_{\rm r}$. Compared to the $g-r$ colour, the $M_{\rm r}$ magnitude is easier to predict with only halo properties. The $R^2$ score is found to be 0.9, and the overall probability distribution of $M_{\rm r}$ is largely reproduced. The top five important features are $\vmax$, $M_{\rm h}$, $\vpeak$, $a_{\rm last}$, and $a_{\rm first}$, with feature importance 0.75, 0.1, 0.03, 0.02, and 0.02, respectively. This suggests that the correlation between magnitude and halo properties is tighter than that between colour and halo properties. Our result is in agreement with \citet{Xu2021b}, who find that halo occupation of a stellar-mass threshold sample of galaxies can be well predicted by only halo properties. Adding galaxy properties to the input features improves the $R^2$ score to 0.98. This improvement is small compared to the $R^2$ improvement of predicting colour when adding galaxy properties (which increases from 0.7 to 0.9).
The top five important features in this prediction are $M_{\rm *}$, $M_{\rm h}$, $M_{\rm coldgas}$, $M_{\rm BH}$, and $a_{\rm last}$, with feature importance 0.92, 0.03, 0.02, 0.01, and 0.01, respectively. These results demonstrate that while baryonic processes may affect central galaxy magnitude, the $M_{\rm r}$ magnitude can be recovered well with only the halo properties.

We then take a further step to investigate the joint distribution of colour and magnitude from the RF model based on only halo properties. Fig.~\ref{fig:ccmd_pred} shows this 2D distribution for fixed halo mass bins. Similar to Fig.~\ref{fig:ccmd_sam}, the black dots show a random fraction of predicted central colour and magnitude for individual haloes, and the solid colour contours show the CCMD components found by GMM. The two CCMD components are obvious in all halo mass bins except for the first one. The original GMM from the SAM are over-plotted as the dashed colour contours for comparison. The RF predicted blue components are much less extended in the colour direction than those in the SAM, due to the lack of predictions of extreme blue centrals. The red components also shrink in both colour and magnitude direction compared to SAM, since the colour and magnitude in the SAM are not fully recovered by the RF prediction with halo properties. Adding galaxy properties to the RF input improves the predicted CCMD components such that they can also be found in the first mass bin, and the blue components are more extended in both colour and magnitude directions, becoming more similar to those in the SAM. However, the blue centrals below $g-r<0.5$ for ${\rm log} M_{\rm h}>12.6$ are still missing. This result demonstrates that both the colour bimodality and the CCMD components can be reproduced to some level based on only halo properties, and the galaxy properties such as black hole mass and cold gas mass are useful for a better recovery. 

\section{CCMD in the Illustris and TNG Simulations}
\label{sec:hydro_ccmd}

\begin{figure*}
	\centering
	\begin{subfigure}[h]{0.8\textwidth}
	\includegraphics[width=\textwidth]{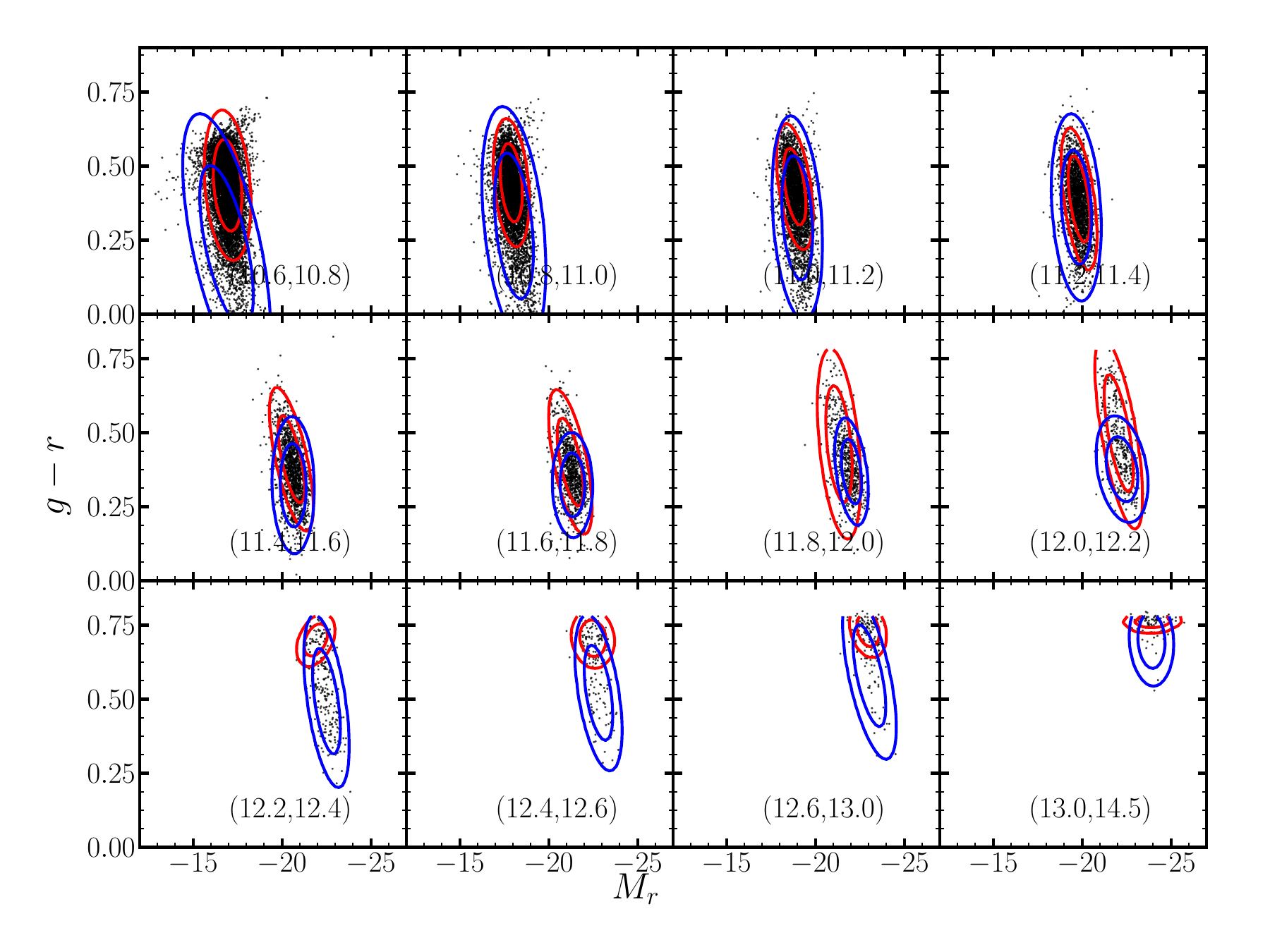}
	\end{subfigure}
	\begin{subfigure}[h]{0.8\textwidth}
	\includegraphics[width=\textwidth]{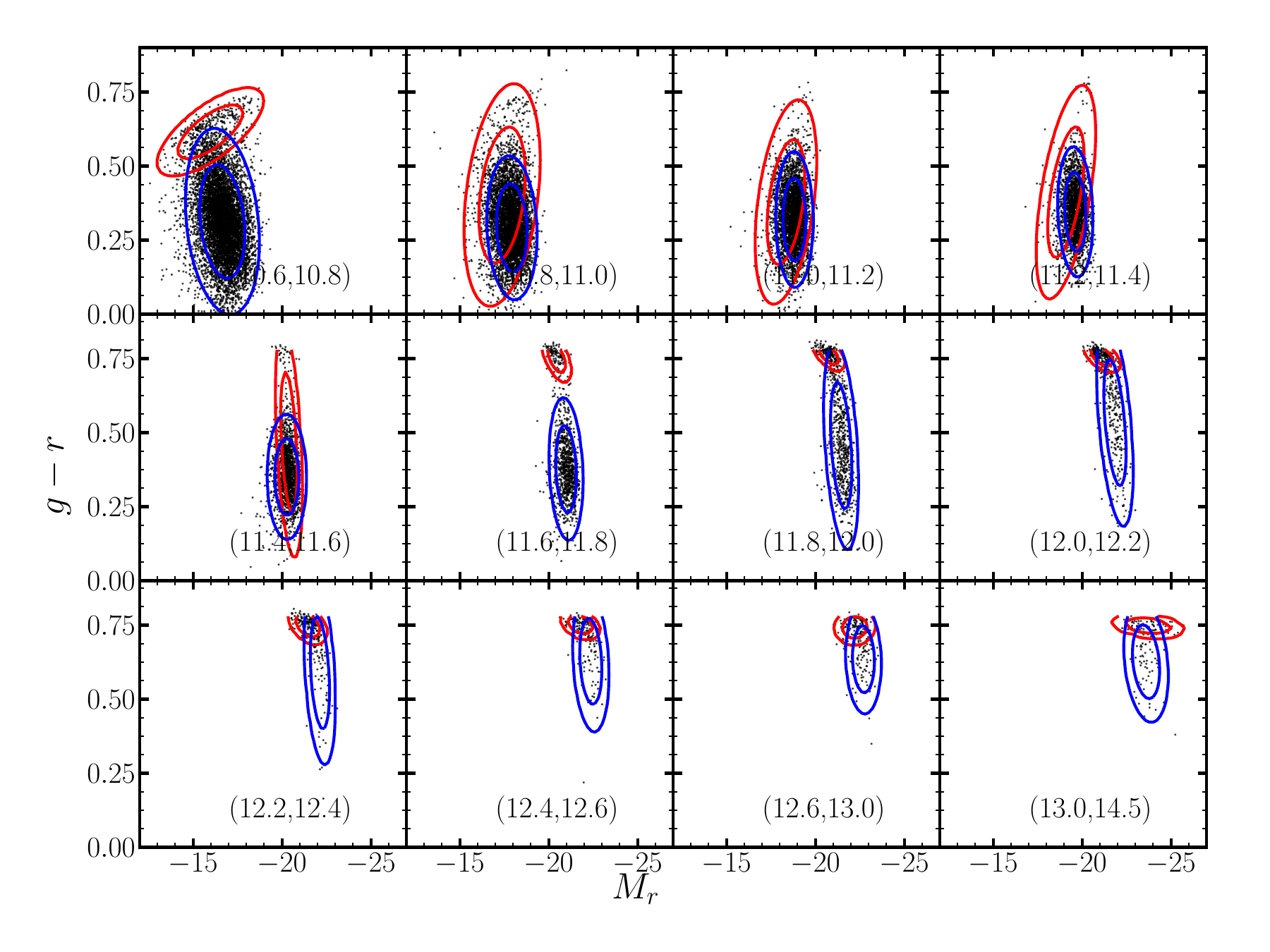}
	\end{subfigure}
\caption{Similar to Fig.~\ref{fig:ccmd_sam}, but showing the CCMD from the Illustris simulation (upper panels) and the TNG100 simulation (lower panels). }
\label{fig:ccmd_illustris}
\end{figure*}

In section~\ref{sec:samCCMD} and section~\ref{sec:RF}, we investigate the central galaxy CCMD and colour dependence on halo and galaxy properties in the SAM. Now we turn to the hydrodynamic simulations. We first perform GMM analysis to find CCMD components in the Illustris simulation and show the results in the upper panels of Fig.~\ref{fig:ccmd_illustris}. For halo mass in the range of ${\rm log} M_{\rm h}<12.2$, the two components are neither distinguishable nor orthogonal. The red sequences are not distinct not only in the GMM but also in the true distributions (dots) of colour and magnitude. The red and blue components can only be distinguished for haloes of ${\rm log} M_{\rm h}>12.2$, and the orthogonality is only seen in the highest mass bin. Notice that we use wider halo mass bins for higher halo mass to ensure a reasonable number of centrals in each bin for the GMM modelling. Compared to the SAM results, the overall colour distribution of central galaxies in the Illustris simulation shifts blue-ward, which leads to slightly bluer means of the blue and red components. Additionally, the red components have a larger spread in colour than those in the SAM at a similar halo mass, suggesting that the Illustris simulation does not produce a tight red sequence. In general, the CCMD components in the Illustris simulation are not as obvious as those in the SAM.

Since the TNG100 simulation is an updated rerun and adopts the same initial conditions as those of the original Illustris simulation, direct comparison between them can reveal the impact of different implementations of physical processes on galaxies. We repeat the GMM analysis for the TNG100 simulation and show the results in the lower panels of Fig.~\ref{fig:ccmd_illustris}. At the red end, the TNG simulation produces galaxies of redder colour than those in the Illustris simulation, resulting in redder red components. In the lowest halo mass bin, distinct orthogonal red and blue components are found. Both the two components are slightly tilted, in the way that the more luminous centrals tend to be redder in the red component and the dimmer centrals tend to be redder in the blue component. Similar to the Illustris, no strong evidence for two CCMD components is seen for haloes in the mass range of $10.8<\log M_{\rm h}<11.6$, mainly because of a rather weak red component (black dots near the top of the distribution). For haloes with $11.6<\log M_{\rm h}<12.2$, the two CCMD components are present clearly again in the TNG simulation but not in the Illustris simulation, a reflection of the updated baryonic physics of the former. The trend of these components in high-mass haloes appears to be opposite to those in low-mass haloes in the tilt direction of the red component such that more luminous centrals are bluer. Interestingly, this trend is consistent with that inferred from the CCMD modelling of SDSS galaxy clustering in \citetalias{XuHaojie2018}. For haloes with $\log M_{\rm h}>12.2$, the red and blue components in TNG are narrower than those in Illustris in colour. Except for the last mass bin, the blue component in TNG is less extended to the blue end than that in Illustris, suggesting that TNG has fewer galaxies with extremely active star formation. 

\citet{Nelson2018} compare colour distribution of the original Illustris and TNG to that of SDSS, and they find that TNG shows a significant improvement in the consistency with SDSS compared to Illustris. They demonstrate that the updated black hole feedback can regulate the colour evolution. Since the CCMD components are also largely identified by colour, the black hole feedback is a key physical process for the origin of the CCMD components.

Comparing the TNG CCMD to the SAM CCMD for the same halo mass bins, namely $11.4<{\rm log} M_{\rm h}<12.6$, the TNG red components are tighter and tilted whereas those in the SAM are essentially horizontal or slightly tilted in the opposite direction. The blue components in the two samples are similar except for a slightly bluer mean colour and narrower magnitude range in the TNG. The mean magnitude of blue and red components in the SAM sample are approximately the same. However, the blue components on average are more luminous than the red components in the TNG. Despite the differences in the details of physical processes implemented and the appearance of CCMD components, the black hole feedback is found to be tightly related to the CCMD components in both SAM and TNG. Finally, the main features and trends in the CCMD from either SAM or hydrodynamic simulations show a broad agreement with those inferred from the CCMD modelling of SDSS galaxy clustering in \citetalias{XuHaojie2018}.

\section{Summary and discussion}
\label{sec:summary}

We investigate the existence and origin of the CCMD components in the SAM and hydrodynamic simulations. For the SAM galaxy sample, we utilise the \citet{Guo2011} model based on the Millennium simulation. We first apply Gaussian mixture modelling to the 2D colour-magnitude distribution of the central galaxies at each fixed halo mass bin. For haloes above ${\rm log} M_{\rm h}=11.4$, the colour-magnitude distribution of central galaxies can be decomposed into two orthogonal 2D Gaussian components, one bluer and the other redder in colour (i.e. the blue and red component). Above ${\rm log} M_{\rm h}=11.8$, the red components tend to occupy a narrow colour range and a wider magnitude range. On the contrary, the blue components are extended in colour and narrow in magnitude. This is in line with the result from \citetalias{XuHaojie2018} based on halo modelling the colour and luminosity dependent clustering of SDSS galaxies within the CCMD framework. 

We then explore the relation between the CCMD components and the host halo properties. Since the red and blue components occupy a similar magnitude range and show a large difference in colour distribution, we focus on the relationship between central galaxy colour and halo properties. We first select two populations with extreme red and blue colours from each halo mass bin and show the differences in their halo properties. Clear correlations between colour and halo assembly properties are seen for haloes of ${\rm log} M_{\rm h}<12.5$. Extreme red centrals are more likely to reside in early-formed haloes with higher concentrations, and extreme blue centrals have the trend to reside in late-formed haloes with lower concentrations. Halo environmental properties of the extreme red and blue populations only show differences at very low halo mass (e.g. ${\rm log} M_{\rm h}<11.6$).

Since galaxy colour may also depend on baryonic processes, we further show its correlation with galaxy properties. In the whole halo mass range, extreme red centrals have higher stellar mass and lower gas specific angular momentum than extreme blue centrals at fixed halo mass. The mass of cold gas and hot gas also show correlations with colour for low-mass haloes, in the way that extreme blue centrals host more cold gas and less hot gas than extreme red centrals. Black hole mass correlates strongly with colour for low-mass haloes. Extreme red centrals host more massive central black holes than extreme blue centrals. However, for haloes with ${\rm log} M_{\rm h}>12.5$, the mass of cold gas, hot gas, or black hole appears to be very similar in extreme red and blue galaxies. 

In addition to the traditional analysis of the relations between colour and halo or galaxy properties, we also perform machine learning studies to further explore these relations. With the SAM sample, we first train a random forest (RF) model with the input features of halo properties to predict galaxy colour and obtain a prediction accuracy of $R^2=0.7$. The colour bimodality is reproduced, but the extreme colours can not be recovered. From the colour distribution, the red peak at $g-r\sim0.4$ is over-predicted compared to the SAM. Besides the predicted colour, the RF model also provides the relative importance of the halo properties. Halo circular velocity related properties and formation history related properties are ranked as the most important factors. While the colour prediction is not ideal, the magnitude is relatively easier to predict with halo properties. This illustrates that magnitude correlates more tightly with halo properties than colour. The two CCMD components can be found in the 2D distribution of predicted colour and magnitude, but the shape and range of the red and blue components are different from those found in the SAM.

We then add galaxy properties to the RF model and re-tune the model for prediction. The $R^2$ score of the new prediction is improved to 0.9, and the overall 1D distribution of colour is closer to the SAM than that based on only halo properties with slightly over-predicted red and blue peaks. The top important properties are black hole mass, cold gas mass, halo formation history properties, and specific angular momentum of gas. As the most important property, black hole mass is an indicator of the strength of AGN feedback. This indicates that the AGN feedback may play an important role in galaxy colour. We find that the overall importance of galaxy properties for colour is higher than that of halo properties. 

Since the CCMD components are mostly separated by colour, the halo or galaxy properties correlated the most with colour can point to the origin of the CCMD components. Investigating the origin of CCMD enables us to improve the CCMD model used to jointly model the colour and luminosity dependent clustering of galaxies. The first CCMD model proposed by \citetalias{XuHaojie2018} based purely on halo mass fits the galaxy clustering in fine luminosity-colour bins well. It indicates that the assembly effect may not be necessary for this model. However, in this work, we find that the CCMD components or galaxy colour show dependencies on both halo secondary properties and galaxy properties in low-mass haloes ($\log M_{\rm h}\lesssim 12.5$) in the galaxy formation models we study. This may lead to possible modifications to the original CCMD model. For high-mass haloes ($\log M_{\rm h}\gtrsim 12.5$), no clear halo or galaxy property is identified to be the controlling factor of determining the colour of central galaxies. This may point to a substantial stochasticity contributed by baryonic processes \citep{Matthee17,Genel2019}. Such a stochasticity, while worth in-depth studying and understanding, would dilute any assembly effect on the galaxy-halo connection, a welcoming feature for halo-based modelling of galaxy clustering. Further studies of CCMD modelling, such as adding assembly parameters to the galaxy-halo connection  or extending to large samples at various redshifts (e.g. those from the DESI survey, \citealt{collaboration2016}), could be helpful for gaining a better understanding of galaxy formation physics, testing galaxy formation models, and creating realistic mock galaxy catalogues. 

\section*{Acknowledgements}
We thank Haojie Xu and Idit Zehavi for useful discussions. This work is supported by the National Science Foundation of China (grant No.11833005,11890692,11621303). XX thanks the support of the University of Utah (partially by a fellowship from the Willard L. and Ruth P. Eccles Foundation) where this work is initiated and Case Western Reserve University where part of this work is accomplished. We gratefully acknowledge the support and resources from the Center for High Performance Computing at the University of Utah and the computing center at National Astronomical Observatories, Chinese Academy of Sciences. XX acknowledges the support from Shanghai Post-doctoral Excellence Program (2021231). QG acknowledges the support by the National Natural Science Foundation of China (No.12033008,11988101).

\section*{Data Availability}
The SAM galaxy catalogue used in tihs work can be access at \url{http://gavo.mpa-garching.mpg.de/Millennium/}. Illustris simulation data and TNG can be access at \url{http://www.illustris-project.org}. 


\bibliographystyle{mnras}
\bibliography{ref_file_ADS}

\bsp	
\label{lastpage}
\end{document}